\documentclass[journal]{IEEEtran}

\usepackage{xcolor}
\usepackage{graphicx}
\usepackage{amsmath,amsfonts}
\usepackage{algorithmic}
\usepackage{algorithm}
\usepackage{array}
\usepackage[caption=false,font=normalsize,labelfont=sf,textfont=sf]{subfig}
\usepackage{textcomp}
\usepackage{stfloats}
\usepackage{url}
\usepackage{verbatim}
\usepackage{cite}
\usepackage{ulem}
\usepackage{amssymb}
\usepackage{booktabs, tabularx}
\usepackage[colorlinks=false,pdfborder={0 0 0}]{hyperref}
\usepackage{units}
\usepackage{multirow,array}
\usepackage{balance}
\usepackage{soul}
\usepackage{color}
\usepackage{cleveref}
\usepackage{wrapfig}
\usepackage{booktabs}
\usepackage{pifont}
\usepackage[utf8]{inputenc}
\usepackage{makecell, tablefootnote, longtable}
\usepackage[nolist]{acronym}

\newcommand{\etal}{et al.}
\newcommand{\eg}{e.\,g.,\ }
\newcommand{\ie}{i.\,e.,\ }

\usepackage{soul}
\usepackage{xcolor}
\definecolor{newgreen}{RGB}{34,139,34}

\begin{acronym}
\acro{PMA}{Permanent Magnet Articulography}
\acro{ACC}{Accuracy}
\acro{ANS}{Autonomic Nervous System}
\acro{AR}{Autoregressive}
\acro{ASR}{Automatic Speech Recognition}
\acro{ATS}{Articulation-to-Speech}
\acro{BCI}{Brain Computer Interface}
\acro{BERT}{Bidirectional Encoder Representations from Transformers}
\acro{biLSTM}{bidirectional LSTM}
\acro{BLS}{Broad Learning Network}
\acro{BVP}{Blood Volume Pulse}
\acro{CNN}{Convolutional Neural Network}
\acro{CNS}{Central Nervous System}
\acro{CNS}{Central Nervous System}
\acro{CSP}{Common Spatial Patterns}
\acro{CWT}{Continuous Wavelet Transform}
\acro{DNN}{Deep Neural Network}
\acro{EDA}{Electrodermal Activity}
\acro{ECG}{Electrocardiogram}
\acro{ECoG}{Electrocorticography}
\acro{ECoG}{Electrocorticography}
\acro{EEG}{Electroencephalography}
\acro{EGG}{Electroglottography}
\acro{EMA}{Electromagnetic Articulography}
\acro{EMG}{Electromyography}
\acro{EOG}{Electrooculogram}
\acro{ETS}{EMG-to-Speech}
\acro{EVC}{Emotional Voice Conversion}
\acro{FFT}{Fast Fourier Transform}
\acro{FIR}{Finite Impulse Response}
\acro{fMRI}{functional Magnetic Resonance Imaging}
\acro{fNIRS}{Functional Near Infrared Spectroscopy}
\acro{F0}{Fundamental Frequency}
\acro{FNN}{Feed-forward Neural Network}
\acro{FDNN}{Feed-forward Deep Neural Network}
\acro{GAN}{Generative Adversarial Network}
\acro{GCCA}{Generalized Canonical Correlation Analysis}
\acro{GMM}{Gaussian Mixture Model}
\acro{GRU}{Gated Recurrent Unit}
\acro{GSR}{Galvanic Skin Response}
\acro{HF}{High-Frequency Power}
\acro{HiFi-GAN}{Generative Adversarial Networks for Efficient and High-Fidelity Speech Synthesis}
\acro{HMM}{Hidden Markov Model}
\acro{HNR}{harmonics-to-noise ratio}
\acro{HRV}{Heart Rate Variability}
\acro{IIR}{Infinite Impulse Response}
\acro{kNN}{k-Nearest Neighbors}
\acro{KRR}{Kernel Ridge Regression}
\acro{LDA}{Linear Discriminant Analysis}
\acro{LF}{Low-Frequency Power}
\acro{LFRCC}{Linear Frequency Residual Cepstral Coefficients}
\acro{LG}{Logarithmic Gaussian}
\acro{LLD}{Low-level Descriptor}
\acro{LPC}{Linear Predictive Coding}
\acro{LPCC}{Linear Predictive Cepstral Coefficients}
\acro{LSF}{Line Spectral Frequencies}
\acro{LSTM}{Long Short-Term-Memory}
\acro{MAV}{Mean Amplutide Value}
\acro{MFCC}{Mel-Frequency Cepstral Coefficients}
\acro{MFSC}{Mel Frequency Spectral Coefficients}
\acro{MRI}{Magnetic Resonance Imaging}
\acro{PNS}{Peripheral Nervous System}
\acro{PNN50}{Percentage of Successive RR Intervals that Differ by more than 50 ms}
\acro{PLSR}{Partial Least Square Regression}
\acro{PPG}{Photoplethysmography}
\acro{RASTA-PLP}{Relative Spectral Transform - Perceptual Linear Prediction}
\acro{RESP}{Respiratory}
\acro{RMS}{Root Mean Square}
\acro{RMSSD}{Root Mean Square of Successive RR Interval Differences}
\acro{RNN}{Recurrent Neural Network}
\acro{SCR}{Skin Conductance Response}
\acro{SCL}{Skin Conductance Level}
\acro{SDNN}{Standard Deviation of NN Intervals}
\acro{SER}{Speech Emotion Recognition}
\acro{SFFS}{Sequential Floating Forward Search}
\acro{SNS}{Somatic Nervous System}
\acro{SoE}{Strength of Excitation}
\acro{SPL}{Sound Pressure Level}
\acro{SRM}{Shared Response Model}
\acro{SSI}{Silent Speech Interface}
\acro{SSR}{Silent Speech Recognition}
\acro{STE}{Speech-to-EMG}
\acro{STFT}{Short-Time Fourier Transform}
\acro{ST}{Semitones}
\acro{SVM}{Support Vector Machines}
\acro{TDNN}{Time-Delay Neural Networks}
\acro{TTS}{Text-to-Speech}
\acro{UAR}{Unweighted Average Recall}
\acro{UTI}{Ultra-Sound Tongue Images}
\acro{VAE-GAN}{Variational Auto-Encoding Generative Adversarial Network}
\acro{VAW-GAN}{Variational Auto-Encoding Wasserstein Generative Adversarial Network}
\acro{WL}{Waveform Length}
\acro{XGBoost}{Extreme Gradient Boosting}
\acro{ZCR}{Zero Crossing Rate}

\end{acronym}

\newcommand{\acbr}[1]{%
   \aclu{#1} (\acs{#1})%
}

\usepackage{fancyhdr}
\begin{document}
%
\title{An Introduction to Silent Paralinguistics}
%
%
%

\author{Zhao Ren,
        Simon Pistrosch,
        Buket Coşkun,
        Kevin Scheck,
        Anton Batliner,
        Björn W. Schuller, and
        Tanja Schultz
\thanks{Z. Ren, B. Coskun, K. Scheck, and T. Schultz are with the Cognitive Sytems Lab, Faculty of Mathematics and Computer Science, University of Bremen, Bremen 28359, Germany (e-mail: zren@uni-bremen.de).}
\thanks{S. Pistrosch, A. Batliner, and B. W. Schuller are with CHI -- the Chair of Health Informatics, TUM University Hospital, Munich 81675, Germany -- and the MCML -- Munich Center for Machine Learning, 80333 Munich, Germany.}
\thanks{B. W. Schuller is also with GLAM -- the Group on Language, Audio, \& Music, Imperial College London, United Kindom.}
\thanks{This study is supported by the Deutsche Forschungsgemeinschaft (DFG, German Research Foundation) through the project ``Silent Paralinguistics" with grant number 40301193.}
\thanks{Manuscript received XX XX, 2025; revised XX XX, 20XX.}}

\markboth{Journal of \LaTeX\ Class Files,~Vol.~XX, No.~XX, XX~20XX}%
{Shell \MakeLowercase{\textit{\etal}}: Bare Demo of IEEEtran.cls for IEEE Journals}

\maketitle
\thispagestyle{fancy}         
\fancyhead{}                     
\lhead{\scriptsize This work has been submitted to the IEEE for possible publication. Copyright may be transferred without notice, after which this version may no longer be accessible.}       \chead{}
\lfoot{}
\cfoot{\thepage}  
\rfoot{}
\renewcommand{\headrulewidth}{0pt}  
\renewcommand{\footrulewidth}{0pt}
\pagestyle{empty}

\begin{abstract}
The ability to speak is an inherent part of human nature and fundamental to our existence as a social species. Unfortunately, this ability can be restricted in certain situations, such as for individuals who have lost their voice or in environments where speaking aloud is unsuitable. Additionally, some people may prefer not to speak audibly due to privacy concerns. For such cases, silent speech interfaces have been proposed, which focus on processing biosignals corresponding to silently produced speech. These interfaces enable synthesis of audible speech from biosignals that are produced when speaking silently and recognition aka decoding of biosignals into text that corresponds to the silently produced speech. While recognition and synthesis of silent speech has been a prominent focus in many research studies, there is a significant gap in deriving paralinguistic information such as affective states from silent speech. To fill this gap, we propose \textit{Silent Paralinguistics}, aiming to predict paralinguistic information from silent speech and ultimately integrate it into the reconstructed audible voice for natural communication.
This survey provides a comprehensive look at methods, research strategies, and objectives within the emerging field of silent paralinguistics.

\end{abstract}


\IEEEpeerreviewmaketitle


\section{Introduction}\label{sec:intro}
Speech is a natural human ability and a core part of what makes us a social species. Unfortunately, the loss of speech ability is a serious concern for many individuals. For instance, in laryngeal cancer therapy the vocal box may be surgically removed during laryngectomy, leaving patients unable to speak audibly. Speech ability can also be significantly degraded due to speech disorders, such as vocal cord paralysis resulting from neurological impairment affecting movement control, as well as spasmodic or functional dysphonia. Apart from speech impairments, individuals may encounter difficulties in certain situations. For example, they may want to avoid discussing private issues in public places, or they struggle to communicate in extremely noisy environments since they lack a strong voice. 

To tackle these challenges, \textit{\acp{SSI}} have been established to enable spoken communication even when the acoustic signal is severely degraded or not available~\cite{denby2010silent}.
\acp{SSI} aim to reconstruct the speech of silent speakers who otherwise are unable or unwilling to speak audibly, via indirect or direct methods by using speech-related biosignals beyond acoustics~\cite{Schultz2017}. The \textit{indirect} method involves recognizing speech content, \ie converting silent speech into text, to then apply text-to-speech methods for synthesizing audible speech. The \textit{direct} method generates audible speech directly from speech-related biosignals.
These biosignals result from the speech production process itself and encompass signals from the brain, neural pathways, articulatory muscles, and articulatory movements. 
Existing research studies have investigated a wide range of speech-related biosignals, including 
electroencephalography (EEG) to convert speech production-related neural signals into text~\cite{herff2015brain} and speech~\cite{Angrick2020}, electromyography (EMG) to convert the activity of articulatory muscles into text~\cite{jou2006towards, wand2011session} and speech~\cite{diener2015direct}, electro-magnetic articulography to trace articulators\cite{9049060}, and ultrasound to capture tongue movements\cite{hueber2011statistical}, to name a few. For a comprehensive overview of the various biosignals, research approaches, studies, and objectives for biosignal-based spoken communication, see~\cite{denby2010silent}, \cite{Schultz2017}, and~\cite{gonzalez-lopez2020}. 

In addition to recognizing text and synthesizing speech from biosignals through SSIs, conveying paralinguistics remains a substantial technical and theoretical challenge. Speech signals are complex and inherently convey various paralinguistic information, such as age, gender, and the emotional state of the speaker. Perceiving such paralinguistic information from speech has been sufficiently investigated in Computational Paralinguistics (CP)~\cite{schuller2014computational,schuller2013}, which aims to automatically recognize paralinguistic information from audible speech through signal processing and Machine Learning (ML) approaches. In contrast, exploring paralinguistic information hidden in biosignals is a research gap in the area of SSIs, both for evaluating the existing SSI systems and for developing novel ML models for synthesizing speech with paralinguistic cues.

To this end, we propose the novel research field of \textit{Silent Paralinguistics (SP)} and highlight its importance in this work. Different from SSIs, SPs focus on extracting not only the textual representation of speech content, but also the paralinguistic information from biosignals during silent articulation. This enables a deeper understanding of silent speech communication through additional paralinguistic cues. Furthermore, the detected paralinguistic cues can be integrated into SSIs to synthesize more natural and expressive speech, enabling more natural speech conversations. 
With this paper, we aim to outline key methods and relevant research strategies in SP, shedding light on the emerging field of SP. We hope to provide valuable insights to the research community and further advance the integration of speech and biosignal processing technologies.


\subsection{Overview and Comparison with Existing Research}
To our knowledge, there is no published work on the topic of SP. Consequently, there is no survey or study introducing the ideas, components, and methods of SP. Nor is there any work relating SP to \acp{SSI}, CP, speech/biosignal-based speech recognition, and speech synthesis / voice conversion. Table \ref{tab:surveys} summarized existing surveys, which mostly treat the above topics separately, rather than highlighting their close integration to achieve natural silent speech communication and their various applications. 

\begin{table}[]
    \centering
        \caption{Survey comparison -- SSIs: Silent Speech Interfaces, \\ CP: Computational Paralinguistics, SR: Speech Recognition, VC: Voice Conversion / Speech Synthesis. }
    \begin{tabular}{lccccc}
    \toprule
         \textbf{Surveys}& \textbf{Year} & \textbf{SSIs} & \textbf{CP} &\textbf{SR} & \textbf{VC}\\
    \midrule
    Denby et al.~\cite{denby2010silent} & 2010 &\checkmark &\ding{55} & \checkmark &\ding{55}   \\
    Schuller and Batliner\cite{schuller2014computational} & 2014  & \ding{55}  &  \checkmark  &  \ding{55}  &  \ding{55}  \\
        Schuller \etal\cite{SCHULLER2015}  & 2015 & \ding{55} & \checkmark & \ding{55} & \ding{55} \\
        Padmanabhan \etal\cite{padmanabhan2015machine} & 2015 & \ding{55} & \ding{55} & \checkmark & \ding{55} \\
        Schultz \etal\cite{Schultz2017}  & 2017 & \checkmark & \ding{55} & \checkmark & \checkmark \\  
        Shu \etal\cite{Shu2018} & 2018 & \ding{55} & \checkmark & \ding{55} & \ding{55} \\
        Bota \etal\cite{Bota2019} & 2019 &  \ding{55} & \checkmark &\ding{55} &  \ding{55} \\
        Gonzalez-Lopez \etal\cite{gonzalez-lopez2020}  & 2020 & \checkmark & \ding{55} & \ding{55} & \ding{55} \\  
        Sisman \etal\cite{sisman2020overview} & 2020 & \ding{55} & \ding{55} & \ding{55} & \checkmark \\          
        Malik \etal\cite{malik2021automatic} & 2021 &  \ding{55} &\ding{55} & \checkmark & \ding{55} \\
        Lee \etal\cite{lee2021biosignal} & 2021 &  \ding{55} &\ding{55} & \checkmark & \ding{55} \\
        Dzedzickis \etal\cite{dzedzickis2020human} & 2022 &  \ding{55} & \checkmark &\ding{55} &  \ding{55} \\
        Ahmad \etal\cite{Ahmad2022} & 2022 & \ding{55} & \checkmark & \ding{55} &  \ding{55} \\
        Zhou \etal\cite{Zhou2022} & 2022 & \ding{55} & \ding{55} & \ding{55} & \checkmark \\
        Yang \etal\cite{yang2022overview} & 2022 & \ding{55} & \ding{55} & \ding{55} & \checkmark \\   
        Prabhavalkar \etal\cite{prabhavalkar2023end} & 2023 & \ding{55} & \ding{55} & \checkmark & \ding{55}  \\
        Kheddar \etal\cite{kheddar2024automatic}& 2024 & \ding{55} & \ding{55} & \checkmark & \ding{55}  \\
       Khare \etal\cite{Khare2024}  & 2024 & \ding{55} & \checkmark & \ding{55} &  \ding{55} \\
    \midrule
        \textbf{This work} & 2025 & \checkmark & \checkmark & \checkmark & \checkmark \\
    \bottomrule
    \end{tabular}
    \label{tab:surveys}
\end{table}

\section{Silent Paralinguistics}
\label{sec:silentpara}

\subsection{Definition of Silent Paralinguistics}
\begin{figure*}
    \centering
    \includegraphics[width=\linewidth]{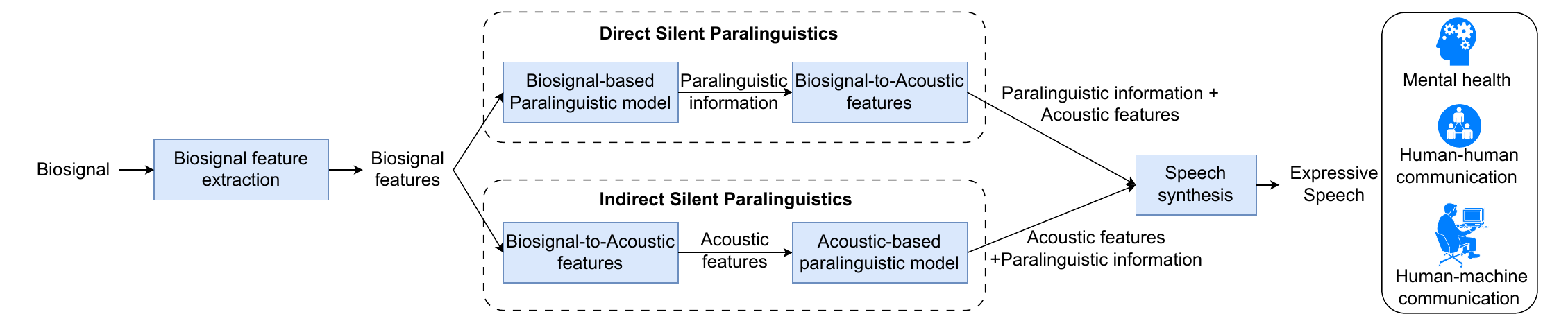}
    \caption{A pipeline of SP. The biosignal features are firstly extracted and fed into the models of either direct SP or indirect SP. The direct SP detects the paralinguistic information and then integrates it into the acoustic features generated by the model of biosignal-to-acoustic feature conversion. The indirect SP generates acoustic features from biosignals, and then recognizes paralinguistic information from the generated acoustic features. Finally, the acoustic features and the paralinguistic information are used to synthesize expressive speech. The applications of SP include mental health improvement, and natural human-human and human-machine communications with paralinguistic cues.}
    \label{fig:sp-framework}
\end{figure*}

To enable natural silent communication, SP is defined as a technology that can perceive, understand, and express paralinguistic information based on speech-related biosignals. Paralinguistic information includes long-term traits (\eg age, gender, and personality), medium-term between traits and states (\eg sleepiness and positive/negative attitude), and short-term states (\eg emotions and emotion-related states)~\cite{schuller2014computational}. 
To create such intelligent machines, SP can be categorized into direct SP and indirect SP (see Fig.~\ref{fig:sp-framework}). The biosignal features are extracted from biosignals, and then used to generate acoustic features and detect paralinguistic information from either direct SP or indirect SP. \textit{Direct SP} includes a set of ML tasks, including perceiving paralinguistic information from biosignals during silent articulation, predicting speech from biosignals (\ie biosignal-to-acoustic feature conversion), and integrating the paralinguistic information into speech for better silent communication. In past research studies, there was limited research investigating paralinguistic information contained in speech converted from biosignals. \textit{Indirect SP} firstly generates acoustic features from biosignals, and then trains an acoustic-based paralinguistic model. Finally, the acoustic features and paralinguistic information are fed into a speech synthesis model to synthesize expressive speech. In addition to speech synthesis, we also introduce speech recognition tasks from biosignals and acoustic speech, which are useful for silent human-machine communication using texts.

\subsection{Applications of Silent Paralinguistics}
\acp{SSI} can enable patients and healthy individuals to communicate with humans and machines using silent speech.
In comparison, SP has potential in a wide range of application areas beyond those of \acp{SSI}. First, SP is promising for improving patients' mental health by enhancing voice naturalness with paralinguistic cues in synthesized audible speech. Second, SP can enable natural silent speech communication with paralinguistic information in a face-to-face communication and in a human-to-machine communication. The applications are discussed in the following.

\subsubsection{Pathological Speech Communication}
One obvious potential use case for silent speech technology lies in clinical applications for individuals who can no longer produce normal audible speech. For instance, patients suffering from laryngectomy cannot produce audible speech without vocal cords. Patients suffering from dysarthria caused by disorders like stroke cannot speak using weak muscles~\cite{enderby2013disorders}. Patients with aphasia\cite{damasio1992aphasia} and apraxia\cite{ogar2005apraxia} also struggle to speak. 
For these individuals, understanding and restoring their voice is crucial to support them in their daily communication. Integrating paralinguistic information can enhance the quality of their speech by providing an emotionally expressive voice or annotated texts that align with their personal characteristics and their current affective state.

\subsubsection{Mental Healthcare for Patients} 
Patients suffering from total laryngectomy can feel depressed about their health and anxiety during social interactions~\cite{mertl2018quality}. Speech disorders, such as dysphonia, can be highly related to mental states of stress, anxiety, and depression~\cite{martinez2015measurement}. 
Audible speech synthesis in silent communication can enable patients to speak with sufficient intelligibility, and expressive speech synthesis can further enhance voice naturalness. 
Improved speech quality can reduce patients' anxiety during communication and alleviate depressed mood. SP shows promise in reducing the risk of mental diseases and enhancing patients' life quality.

\subsubsection{Human-Human Private Communication} 
Silent speech technology can enable private conversations that cannot be overheard by surrounding people. For instance, speakers in phone or video calls have to be silent to protect their privacy, especially in public environments~\cite{gaddy2022voicing}. Additionally, it is easier to communicate with the technologies in SP in noisy environments. Furthermore, perceiving paralinguistic information is helpful for understanding personal traits and states from the generated text or speech content. The paralinguistic information can be integrated into the speech, so that the generated voice sounds age and gender appropriate or even exactly the same as the original voice.

\subsubsection{Silent Human-Machine Communication}
Conversational AI using large language models has been nowadays very popular, particularly through the widespread adoption of various advanced chatbots, \eg chatGPT\footnote{\url{https://openai.com/index/chatgpt/}}, DeepSeek\footnote{\url{https://www.deepseek.com/}}, etc. In the meanwhile, voice assistants, such as GPT, allow humans to interact with a machine using voice. For silent communication, \acp{SSI} can create an interface between biosignals captured during silent articulation and these conversational assistants. Either generated texts or audible speech can be the input of conversational assistants. More importantly, it is essential to perceive paralinguistic information for an AI assistant to better understand a user. Enhanced input, \ie texts annotated with traits/states and expressive audible speech, will provide more sufficient information to an AI assistant. 

\section{Biosignals of Speech Production}\label{sec:bio}
Speech production begins in the brain, then activates the muscles, leading to respiratory, laryngeal, and articulatory movements~\cite{Schultz2017}.
To understand speech content and paralinguistic information from biosignals, there are various available speech-related signals obtained from brain, muscle, articulatory, and laryngeal activities~\cite{Schultz2017}.
This section provides a general overview of biosignals which are required for SP. 
The biosignals can be summarized as surface \ac{EEG}, \ac{fNIRS}, \ac{ECoG}, intracortical singals, \ac{EMG}, video images, ultrasound images, \ac{EMA}, \ac{PMA} and \ac{EGG}. Functional Magnetic Resonance Imaging (fMRI) 
could also be relevant; however, due to its size and cost, it is not wearable/mobile and thus excluded from this survey. 

\subsection{Biosignals of Brain Activity} 
\textbf{Surface \acbr{EEG}} measures the electrophysiological activity of the brain using non-invasive electrodes. It detects the electrical fields produced by the activation of neurons. The EEG signal can be collected using multiple electrodes placed on different brain regions according to the standard 10-20 topology~\cite{Li2022}. \ac{EEG} provides a cost-effective and easy-to-use setup for \acp{SSI}\cite{Tabar2017,Fitriah2022}. 
However, \ac{EEG} is highly susceptible to artifacts~\cite{biosignals09,Schultz2017}. Despite this, Porbadnigk \etal~supported the hypothesis that temporal correlated artifacts were recognized instead of words in block recordings in their experiments~\cite{biosignals09}. 
Many recent studies~\cite{Cooney2022,10078589,Ghosh2023,Kim2023,Pawar2023,Pan2023} have employed \ac{EEG}-based \acp{BCI} for \acp{SSI}, and primarily focused on sound or word-level analysis rather than implementing an open-vocabulary speech synthesis system. By combining \ac{EEG} and \ac{EMG}, it has been shown to be promising in synthesizing acoustic speech~\cite{Li2023ssi}.
Furthermore, as \ac{EEG} directly captures the electrical activity of the brain cortex, it can be considered a reliable method to recognize emotional states~\cite{Kamble2023}. \ac{EEG} can record the physiological activity of the central nervous system, which regulates whole-body physical and physiological activities to participate in the emotional process~\cite{Li2022,edgar2020}. Schaff and Schultz worked on emotion recognition from EEG signals in an early study~\cite{schaaff2009towards}, and researchers have conducted numerous recent studies using \ac{EEG} in the area of emotion recognition~\cite{Fayaz2024, Mutawa2024, Khan2024, Avola2024, Saha2024, Wang2024EEG, Khubani2024, YiwuWang2024, Fu2024, Fan2024, Patel2024}.

\textbf{\acbr{fNIRS}}~\cite{jobsis1977noninvasive} is a non-invasive brain imaging method that measures blood oxygenation levels due to neuron activation, which involves the transfer of ions across cell membranes powered by increased oxygen metabolism. Infrared light is absorbed by hemoglobin but not by biological tissues; therefore, by detecting the amount of light absorbed, hemoglobin levels can be determined. The fNIRS is less susceptible to noise compared to \ac{EEG}~\cite{Schultz2017}; yet, its depth sensitivity is only about 0.75 mm below the brain surface, as infrared light has limited penetration into cerebral tissue~\cite{gonzalez-lopez2020}. More importantly, \ac{fNIRS} signals can exhibit a delay due to the time required for blood to flow to the measurement site in response to neural activity. The \ac{fNIRS} provides a semi-portable setup for \acp{SSI}, allowing participants to wear a device with light emitters and detectors and communicate in a comfortable environment~\cite{herff2016automatic}. Herff \etal~\cite{herff2013self} took a first step toward detecting speech activity from fNIRS singals.
Liu and Ayaz~\cite{liu2018speech} recognized speech using \ac{fNIRS} signals. 
In~\cite{Cooney2022}, \ac{fNIRS} and \ac{EEG} were employed together to decode imagined speech. Studies in~\cite{heger2014continuous,Chen2024_fNIRS,Jin2023} also use \ac{fNIRS} for emotion recognition. Putze \etal~\cite{putze2014hybrid} investigated the classification of auditory and visual perception from fNIRS and EEG signals.

\textbf{\acbr{ECoG}} is an invasive method that measures the electrical activity of the brain directly from the cortical surface. ECoG is mostly used as part of epilepsy surgeries~\cite{worrell2008electrocorticography}. However, although it is an invasive and impractical method, studies have been published using ECoG-based \acp{SSI}. Herff \etal~\cite{herff2015brain,herff2016automatic}, for the first time, decoded continuously spoken speech into expressed words from \ac{ECoG} signals, using their proposed Brain-to-Text system. In~\cite{Meng2023}, participants with already temporarily implanted ECoG for intractable epilepsy have been investigated.
In~\cite{9049060}, \ac{ECoG} is also used for emotion recognition; however, this is not suitable for SP due to its invasive nature. 

\textbf{Intracortical recording} captures intracortical signals with Utah or NeuroPort neurotrophic microelectrodes implanted in the cerebral cortex of the brain through surgeries~\cite{Schultz2017,brumberg2011classification}.
Intracortical signals have been employed for imagined speech decoding, including intended phoneme prediction~\cite{brumberg2011classification} and speech synthesis~\cite{guenther2009wireless,brumberg2010brain}. A few studies have investigated the relationship between intracortical signals and personal states, such as emotion and mood~\cite{quettier2024individual}, and decoded mood variations from intracortical signals~\cite{sani2018mood}. However, intracortical signals are not suitable for SP as the implantation procedure is invasive.

\subsection{Biosignals of Muscle Activity} \label{sec:bio-muscle}
\textbf{\acbr{EMG}} measures the electrical activity produced by muscle contractions using either surface or intramuscular electrodes~\cite{dzedzickis2020human}. The properties of surface EMG signals are influenced by the characteristics of the tissue that separates activated muscles from the electrodes, as well as by the physiological and anatomical properties of muscle fibers, including their numbers, positions, and orientations~\cite{gonzalez-lopez2020}. In the field of \acp{SSI}, surface electrodes are preferred to capture facial muscle activities, due to their non-invasive nature compared to intramuscular electrodes~\cite{Schultz2017,gonzalez-lopez2020}. We will mainly introduce surface EMG signals in this work, namely EMG.

Facial EMG signals were collected and processed to recognize speech in early studies of Schultz and Wand \etal~\cite{schultz2010modeling,wand2011session,wand2014tackling}. As EMG-to-text-to-speech is an indirect way to synthesize speech, recent studies directly synthesize acoustic features from facial EMG and use a vocoder and reconstruct acoustic speech~\cite{diener2016initial,janke2017emg,gaddy2020digital,gaddy2021improved,Scheck2023b}. 
Moreover, facial \ac{EMG} has been commonly utilized in emotion recognition because emotions are intricately linked to actions and facial muscles~\cite{ekman1982methods}. Gibert \etal~\cite{gibert2009enhancement} validated the feasibility of detecting emotion from facial EMG signals. Veldanda \etal~\cite{veldanda2024can} investigated facial action unit recognition based on facial EMG signals.
There are also studies~\cite{Kose2021, Manju2022, Sato2022} employing the trapezius muscle in the upper back. 
According to the study in \cite{li2009semg}, the trapezius muscle is also a component of the emotional motor system, capable of activation without voluntary contraction or physical load, similar to facial muscles. However, the trapezius muscle cannot replace facial muscles to capture speech-relevant muscle movements in the context of \acp{SSI} and SP.
Various studies have been conducted in the area of emotion recognition using facial \ac{EMG} and trapezius \ac{EMG}~\cite{Kose2021, Manju2022, Bhatlawande2024, Qin2024, Li2023, Han2023, Khan2023, Branco2022, Chen2022}.

\subsection{Biosignals of Articulatory Movement}
\textbf{\acbr{EMA}} or \textbf{\acbr{PMA}} uses magnetic tracers to detect how articulators move. The setup for \ac{EMA} involves receiver coils attached to key articulators such as the lips, tongue, and velum. These coils are connected via wires to external equipment that records the voltage induced by changes in the surrounding magnetic field, allowing the system to record articulatory movements. In PMA, permanent magnet transmitters are mounted on the articulators, and an external sensor measures the aggregated magnetic field. \ac{EMA} is invasive and requires wires to be run inside the mouth~\cite{Schultz2017}. Instead, \ac{PMA} is more comfortable as it only applies magnets to be fixed inside the mouth without connecting wires~\cite{Schultz2017,gonzalez-lopez2020}.
As EMA and PMA provide a close reproduction of the articulatory movement, they were extensively researched for speech recognition and speech synthesis~\cite{livescu2005feature,ling2009integrating,Schultz2017}.
\ac{EMA} or \ac{PMA} is not a common approach for detecting paralinguistic information; however, several studies have focused on \ac{F0} detection using \ac{EMA}~\cite{Liu2016,Zhao2018,9053231}.

\textbf{\acbr{EGG}} measures the electrical impedance of the glottis using electrodes placed on the throat, which helps track vocal cord movement. When the vocal cords are closed, their contact area peaks, resulting in the lowest resistance and the highest recorded voltage in the EGG. Conversely, when the vocal cords are open, the recorded voltage is at its lowest value~\cite{Chen2022}. To the authors' knowledge, although \ac{EGG} has been used for speech recognition and speech synthesis~\cite{Chen2022egg}, it has not been used for SSIs. It has been shown effective
in paralinguistics studies~\cite{schuller2014interspeech,Liu2023, Grigorev2022, Hui2015,Chen2013}.

\textbf{Video and Ultrasound Imaging} are often used together for silent speech synthesis~\cite{HUEBER2010288,hueber2011statistical}. Video imaging serves as a straightforward and effective method to detect movements of external articulators like the lips and jaw. Based on video images, lipreading has been explored to recognize vowels in~\cite{watanabe1990lip}.
Various lip-to-speech systems were further proposed in recent years~\cite{ephrat2017lip2speech,  Choi23Lip2Speech, Hedge23Lip2Speech, Dong24Lip2Speech, zheng2024lip2speech, Kim2024Lip2Speech, Prajwal2020CVPRLip2Speech}. Facial images have been widely used to predict paralinguistic information, such as personality traits~\cite{xu2021prediction}, sleepiness~\cite{vargas2017facial}, and emotional states~\cite{li2020deep}.

Ultrasound imaging, on the other hand, primarily focuses on capturing tongue movements within the vocal tract during speech. Ultrasound imaging has been employed to recognize and synthesize speech for \ac{SSI}~\cite{csapo2017dnn,toth2018multi,kimura2019sottovoce,T_th_2023}. 
However, to the best of the authors' knowledge, ultrasound imaging has not yet been utilized for emotion recognition. Some studies~\cite{Grosz2018,Dai2021} focused on \ac{F0} detection using ultrasound. Recently, several studies~\cite{Su696,gao2020echowhisper} have also focused on \acp{SSI} using ultrasound recorded by smartphones.

\section{Speech Recognition and Synthesis}\label{sec:SSI}
As aforementioned in Section~\ref{sec:silentpara}, recognizing speech is helpful for silent communication between human and conversational assistants. In this section, we will introduce automatic speech recognition which can be used to further analyze generated acoustic features in the indirect SP in Section~\ref{sec:asr}, and then we discuss
biosignal-based speech recognition in Section~\ref{sec:biosingal-recognition-synthesis}. In Section~\ref{sec:biosingal-recognition-synthesis}, speech synthesis from biosignals will be also presented as a part of SP.

\subsection{Automatic Speech Recognition}\label{sec:asr}
\ac{ASR} aims to recognize spoken words from audible speech, \ie converting spoken speech into text form~\cite{malik2021automatic,kheddar2024automatic}. Given a speech form as input, ASR includes pre-processing, feature extraction, an acoustic model, and a language model.

\subsubsection{Pre-Processing} In the real world, speech is often spoken in noisy environments. The pre-processing module reduces noise associated with speech using several potential methods, such as windowing, normalization, end-point detection, and pre-emphasis. 

\subsubsection{Feature Extraction} The feature extraction module mainly relies on frequency-based features, such as \acp{MFCC}, linear predictive coding, and discrete wavelet transform. More recently, it has been shown that feature extraction can be combined with the acoustic model. Therefore, raw speech signals are used as input, forming end-to-end \ac{ASR}~\cite{ao2022pre}.

\subsubsection{Acoustic Model} An acoustic model is trained to predict phonemes by taking the extracted features as input. An acoustic model usually employs \acp{HMM} and \acp{FNN} to form a classification task~\cite{trentin2001survey}. The probabilities of phonemes are predicted by an acoustic model based on speech features. 
In the past decades, \ac{ASR} techniques mainly benefited from deep learning (DL) models, such as \ac{LSTM}-RNNs, and their combination with CNNs~\cite{passricha2019hybrid}. More recently, transformers have demonstrated their effectiveness in \ac{ASR} due to their complex structures and their capability of taking sequential information into account~\cite{zeyer2019comparison}. 

\subsubsection{Language Model} In early \ac{ASR} systems, the 
pronunciation dictionary played a central role, often relying heavily on manually crafted phoneme-to-word mappings~\cite{rabiner2002tutorial}. The predicted phonemes are mapped to words using a pronunciation dictionary. Nowadays, a language model recognizes the phoneme predicted by the acoustic model, and generates $n$-grams,
words, and sentences based on predicted phonemes~\cite{malik2021automatic}. Language models can be categorized into static and dynamic models. A typical static model is an $n$-gram, which forms a sequence of $n$ adjacent symbols, \eg letters and words. An $n$-gram predicts the occurrence probability of a symbol occurring, given the preceding $n-1$ symbols. However, $n$-grams are unable to process symbol dependencies longer than $n$~\cite{troncoso2005trigger}. In contrast, dynamic models have a broader scope with longer symbol dependencies using various methods, such as triggers~\cite{troncoso2005trigger}. Neural network-based language models in natural language processing have been popular in recent studies. RNNs ~\cite{hori2018end} and transformers have been successfully used as language models~\cite{baquero2020improved}.

\subsection{Biosignal-based Speech Recognition and Synthesis}\label{sec:biosingal-recognition-synthesis}
\begin{figure}[ht]
    \centering
    \includegraphics[width=0.45\textwidth]{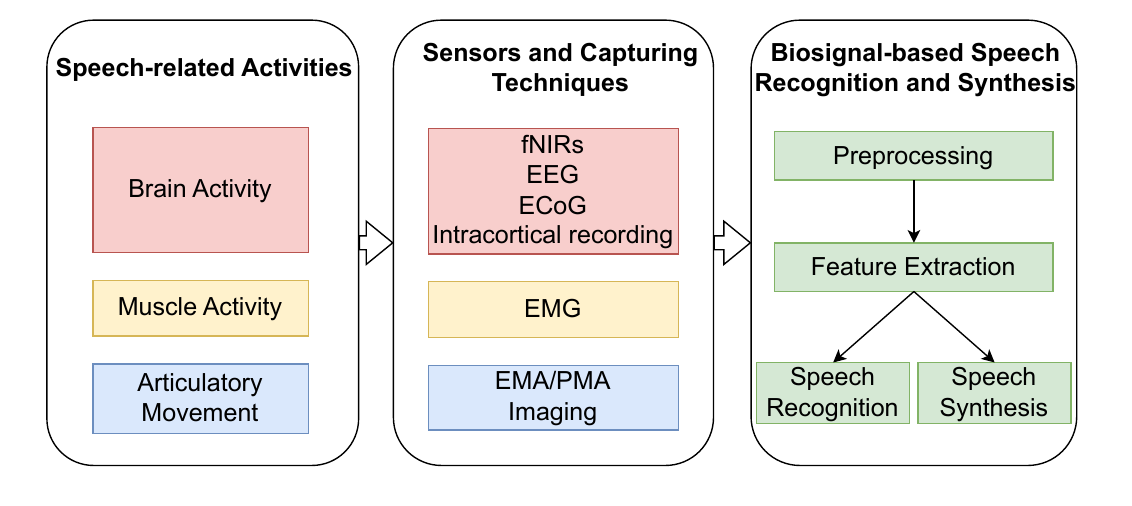}
    \caption{The pipeline of speech recognition and synthesis from biosignals. 
    }
    \label{fig:ssi}
\end{figure}

\acp{SSI} can recognize speech and synthesize acoustic speech signals from biosignals which are emitted typically during silent speech production processes~\cite{gonzalez-lopez2020}. 
Various biosignals discussed in Section~\ref{sec:bio} can be used for speech recognition and synthesis. 
As shown in Fig.~\ref{fig:ssi}, after biosignals are acquired, preprocessing, feature extraction, speech recognition, and speech synthesis procedures are conducted. Particularly, in speech synthesis, audible speech can be firstly recognized as texts and then the texts are synthesized as audio using a \ac{TTS} system. Alternatively, audible speech can be synthesized directly from biosignals.

\subsubsection{Pre-Processing}
Biosignals are susceptible to external artifacts and noise which need to be removed. 

\textbf{Biosignals of Brain Activity.}
A bandpass filter is generally employed to remove artifacts from EEG, ECoG, and intracortical signals, whereas a notch filter is specifically employed to remove powerline noise. Standard filters are Butterworth~\cite{butterworth1930theory,wilson2020decoding}, finite
impulse response, and  infinite impulse response. Also, baseline removal is mostly used for preprocessing. 
Additionally, independent component analysis has been utilized for noise removal from \ac{EEG}~\cite{NIPS1995_754dda4b}. 
In \ac{fNIRS} studies, the modified Beer-Lambert Law~\cite{cope1988system} is mostly used as an initial step to convert light intensity into oxygenated-hemoglobin and deoxygenated-hemoglobin concentrations. 

\textbf{Biosignals of Muscle Activity.}
Similar to brain signals, bandpass filters are often employed to remove noise from \ac{EMG} signals. Movement artifacts are often at low frequencies, whereas random noises have high frequencies~\cite{Lai2023}. Therefore, bandpass filters are usually set between 15-28\,Hz and 400-450\,Hz.
In~\cite{Lai2023}, a wavelet denoising method was used for \ac{EMG} signals, as wavelet transformation has good frequency resolution at high frequencies. 

\textbf{Biosignals of Articulatory Movement.}
To recognize and synthesize speech, low pass filtering is often applied to remove noise from \ac{EMA} and PMA signals~\cite{wang2014preliminary}. Due to anatomical differences in speakers' tongues, normalization of articulatory movements is essential to ensure the generalizability of EMA-based and PMA-based models across speakers~\cite{GILBERT20101189,wang2014preliminary}.
Video imaging data is typically obtained in high-resolution 2D or 3D formats and then processed further as images. Transformation is often used to augment the training set using various methods, such as scaling, rotating, etc.~\cite{wrench2022beyond}. In \cite{beeson23_interspeech}, raw ultrasound scanline data was converted into videos for further process using UltraSuite Tools~\cite{eshky18_interspeech}.

\subsubsection{Feature Extraction}
Various feature extraction methods have been applied to biosignals, depending on the signal types.

\textbf{Biosignals of Brain Activity.}
In \ac{SSI} systems using \ac{EEG} and \ac{ECoG}, time-domain features (such as mean, median, and standard deviation) and temporal features are employed, as well as frequency-domain features, dividing 
into alpha, beta, gamma, high gamma, delta, and theta bands. Frequency domain features are extracted as time-frequency representations, including wavelet coefficients~\cite{57199}, \ac{MFCC}, \ac{MFSC}, and \ac{LPC}~\cite{Ramkumar2023}.
Particularly, the studies in Herff \etal~\cite{herff2015brain} and Angrick \etal~\cite{Angrick2020} utilized the broadband gamma band (70-170\,Hz) relevant to speech and language for \ac{ECoG} signals. The authors~\cite{herff2015brain} computed the signal energy and then applied a natural logarithm for feature extraction. 
Furthermore, to identify complex information in \ac{EEG}, DL-based feature extraction approaches, such as \ac{CNN} or Conditional Autoencoder, are also employed~\cite{Kim2023}.
For \ac{fNIRS} signals, linear data trend and spatial filters, such as \ac{SRM}~\cite{chen2015reduced} and \ac{GCCA}~\cite{SHEN2014310,liu2018speech}, are commonly used for feature extraction. Regarding feature extraction from intracortical signals, features usually include spike-band powers~\cite{wairagkar2023synthesizing}, spikes, and local field potential powers~\cite{stavisky2018decoding}.

\textbf{Biosignals of Muscle Activity.}
For the \ac{EMG}-based \acp{SSI}, time, frequency, and time-frequency domain features are extracted. Time-domain features include features such as mean, power, and zero-crossing rate~\cite{jou2006towards, hudgins1993new}. For the time-frequency representations, the \ac{STFT} and wavelet transform coefficients~\cite{57199} are commonly preferred~\cite{1566521,deng2014towards}. The raw EMG signals without feature extraction can be also fed into \ac{SSI} systems in recent studies~\cite{Scheck2023,ren2024diff}.  

\textbf{Biosignals of Articulatory Movement.}
\ac{EMA} and \ac{PMA} data can be used directly without a feature extraction process~\cite{Schultz2017}.
For the imaging data, mostly feature extraction is conducted using deep architectures such as \acp{CNN}. 


\subsubsection{Biosignal-based Speech Recognition}
Biosignal-based speech recognition aims to recognize speech content, \ie generate texts from biosignals.

\textbf{Biosignals of Brain Activity.}
The distributions of silent words can be statistically similar, particularly for words related to directed movements, which tend to share similar patterns of brain activity~\cite{vorontsova2021silent}. Early studies focused on classifying isolated phones, syllables, or words from EEG.
For instance, Ghane \etal~\cite{ghane2015robust} applied statistical analysis for decoding EEG into imagined speech phones. The work in~\cite{brigham2010imagined} predicted two imagined syllables using a $k$-nearest neighbour classifier. Gonzalez \etal~\cite{vorontsova2021silent} used \ac{CNN} and \ac{RNN} models to classify nine words from EEG collected during silent speech.
Due to the development of DL, continuous sentences can be generated in recent studies.  In~\cite{krishna2020continuous}, silently spoken sentences were recognized from EEG using a Connectionist Temporal Classification (CTC)-based \ac{ASR} model. At the inference stage, a CTC beam search decoder is combined with a 4-gram language model for generating texts.

Early studies in ECoG focused on recognizing isolated phones and words, 
for example, 
using \ac{SVM}~\cite{blakely2008localization} and a linear classifier~\cite{mugler2014direct}.
Additionally, many studies demonstrated the potential of ECoG to predict continuous sentences. Herff \etal~\cite{herff2015brain}, for the first time, predicted phones using Gaussian models, and then a pronunciation dictionary and an $n$-gram language model were applied to predict sentences.  
Texts were generated from ECoG using an encoder-decoder model with transformers in~\cite{komeiji2024feasibility}. Yuan \etal~\cite{yuan2024improving} improved ECoG-based speech recognition by training a Wav2Vec model on additional unlabeled ECoG for extracting ECoG representations, and the ECoG representations are then fed into an encoder-decoder model for speech recognition.

Studies have investigated \ac{fNIRS} for speech activity, and pointed out \ac{fNIRS} is not suitable for speech recognition~\cite{herff2013self,herff2016automatic}. A few studies recognized isolated words from \ac{fNIRS} using ML models such as logistic regression~\cite{liu2018speech}.

Studies have investigated the classification of phonemes from intracortical signals using classic ML models, such as \acp{SVM}~\cite{stavisky2018decoding} and logistic regression~\cite{wilson2020decoding}. DL models, such as \acp{RNN}, have been found to perform better than classic ML models for phoneme classification~\cite{wilson2020decoding}. 

\textbf{Biosignals of Muscle Activity.}
In early studies, Maier-Hein \etal~\cite{1566521} firstly worked on speech recognition from facial \ac{EMG} signals using an \ac{HMM} model. Jou \etal~\cite{jou2006towards} trained context-independent acoustic models and combined them with a trigram language model to predict sentences.
Further works also used \acp{HMM} and language models to recognize speech from \ac{EMG} signals~\cite{janke2010impact,janke2012further}.
In~\cite{wand2014emg,wand2014tackling}, an \ac{HMM} model and a linguistic model were used to decode texts from EMG. Wand \etal~\cite{wand2014towards} investigated the feasibility of EMG-based speech recognition by training a model on EMG data recorded in multiple sessions and applying it to a new session. 
More recently, several models, including \ac{SVM}, DenseNet, ResNet, \ac{biLSTM}-RNNs, and transformers, were applied to EMG-based speech recognition in~\cite{cha2022deep,li2023semg,song2023decoding}.

\textbf{Biosignals of Articulatory Movement.}
Similar to brain and muscle biosignals, conventional ML models, such as \acp{SVM} and \acp{HMM}, were used to predict isolated word- or sentence probabilities from articulatory movements in early studies~\cite{wang2012sentence,hofe2013small}. In recent years, DL has been shown promising for recognition tasks.
For instance, Cao \etal\cite{Cao2023} utilized \ac{GMM}- and \ac{FDNN}-based approaches for \ac{EMA}-based vowel classification and speech recognition.
Another study~\cite{beeson23_interspeech} employed a typical \ac{FDNN}-\ac{HMM} pipeline for processing tongue ultrasound and lip videos.

\subsubsection{Biosignal-to-Speech Conversion}
Beyond texts, acoustic speech contains a wealth of information, conveying important markers of human communication such as emotion and speaker identity~\cite{herff2020towards}.
Compared to Biosignal-based speech recognition, Biosignal-to-Speech conversion is more efficient for generating audio in real-time than recognizing text first and then synthesizing audio with a \ac{TTS} system. This technique holds the potential for real-time streaming speech synthesis, as audio is generated immediately after the articulation movement when the model latency is low.

\textbf{Biosignals of Brain Activity.}
Based on biosignals of brain activity, acoustic features are firstly predicted using an encoder, and then the acoustic speech is generated by a vocoder, such as a pre-trained HiFi-GAN~\cite{kong2020hifi}.
Generative models have been successfully used as an encoder to predict acoustic features from \ac{EEG} signals~\cite{10078574}.
For instance, Lee \etal~\cite{lee2023towards} reconstructed Mel spectrograms of imagined speech from EEG using \acp{GAN}~\cite{goodfellow2014generative}. In addition to the GAN loss, the generative model was trained with a reconstruction loss by comparing the generated Mel spectrograms and the Mel spectrograms of spoken speech, as well as a CTC loss based on the texts generated by a vocoder and an \ac{ASR} model. 
In another study~\cite{10078574}, 
the generative model included \acp{GRU}~\cite{cho2014learning} and a multi-receptive field fusion to generate Mel spectrograms. In addition to the prediction of Mel spectrograms, a generative model, including \ac{GRU} layers and attention layers, was proposed to predict \acp{MFCC} for EEG-based speech synthesis in~\cite{9441306}.  

Based on \ac{ECoG} signals, a pilot study by Herff \etal~\cite{herff2016towards} applied a Lasso regression to predict speech spectrogram from \ac{ECoG} features, and then reconstructed speech using an algorithm with \ac{STFT} and inverse \ac{STFT}.
Herff \etal~\cite{herff2020towards} also employed Unit Selection, which was developed to process small-scale datasets, to generate speech spectrograms by concatenating speech units. 
Angrick \etal~\cite{Angrick2020} noted that regression tasks can generate unexpected amplitude spikes and unnatural volume increases in synthesized speech, especially when the input data exhibits large variations.
Therefore, Angrick \etal~\cite{Angrick2020} proposed a quantization-based approach to discretize the continuous space of log Mel spectrograms into a manageable number of disjoint intervals, and applied \ac{LDA} as a classification model to estimate log Mel spectrograms of speech. Similarly, Meng \etal~\cite{Meng2022,Meng2023} implemented a closed-loop \ac{BCI} system for real-time speech synthesis from  
the  
\ac{ECoG} signal using discretization and 
an \ac{LDA} classifier-based approach. More recently, due to the strong capability of DL in representation learing, \acp{MFCC} were predicted by an encoder-vocoder model in a regression task~\cite{komeiji2024feasibility}.

Intracortical signals have been applied to synthesize speech in only a limited number of studies because of their invasive nature. The study in~\cite{wairagkar2023synthesizing} trained \acp{RNN} to predict low-dimensional speech features, including pitch period, pitch strength, and cepstral coefficients, and reconstruct speech with an LPCNet vocoder~\cite{valin2019lpcnet}. A causal model was further developed to enable real-time prediction of low-dimensional speech features in~\cite{wairagkar2025instantaneous}.

\textbf{Biosignals of Muscle Activity.}
EMG-to-Speech conversion studies mostly focused on predicting 
spectral features of speech such as \ac{F0}, power~\cite{toth2009synthesizing,diener2015direct}, and Mel spectrograms~\cite{diener2015direct}. The features are then fed into a vocoder to synthesize audible speech~\cite{diener2015direct}. 
To map EMG to spectral features, the initial approaches generally involve \acp{GMM} and \acp{HMM} as the model architectures~\cite{Schultz2017,gonzalez-lopez2020}. Recent approaches use \acp{FDNN} for this mapping~\cite{diener2015direct}. Gaddy and Klein~\cite{gaddy2020digital} proposed an \ac{LSTM}-RNN model for encoding the EMG-Signal to Mel spectrograms. Furthermore, they suggested an improved encoder using transformers and convolutional layers~\cite{gaddy2021improved}. A diffusion model was applied by Ren \etal~\cite{ren2024diff} to enhance the quality of the acoustic features that an EMG encoder predicted, thereby improving speech naturalness. Scheck \etal~\cite{Scheck2023b} suggested a causal model architecture for a low-latency EMG-based speech synthesis approach.
As alternative to acoustic speech features, recent work has also investigated the use of soft-speech units extracted by self-supervised learning (SSL) speech models, such as HuBERT~\cite{Hsu21huBERT}, as training targets~\cite{Scheck2023,scheck2023ste}. 


\textbf{Biosignals of Articulatory Movement.}
Similar to speech synthesis from biosignals of brain and muscle activities, many studies on synthesizing speech from biosignals of articulatory movement also focused on predicted acoustic features.
Cao \etal~\cite{cao2021investigating} and Wu \etal~\cite{wu22i_interspeech} proposed direct \ac{ATS} conversion based on \ac{EMA}. Multiple acoustic features were predicted, including log-\ac{F0}, band aperiodicities, Mel-generalized cepstrals, and voiced/unvoiced labels~\cite{cao2021investigating}.
Gonzalez and Green~\cite{GONZALEZ2018148} proposed an \ac{LSTM}-RNN-based real-time silent speech system that predicted acoustic \ac{MFCC} features from \ac{PMA}.
In~\cite{10096920}, Zheng \etal~reconstructed Mel spectrograms of speech using pseudo-target generation and domain adversarial training from ultrasound tongue images and optical lip videos. 

\section{Computational Paralinguistics from Speech and Biosignals}\label{sec:CP}
\begin{figure*}[ht]
    \centering
    \includegraphics[width=1.0\textwidth]{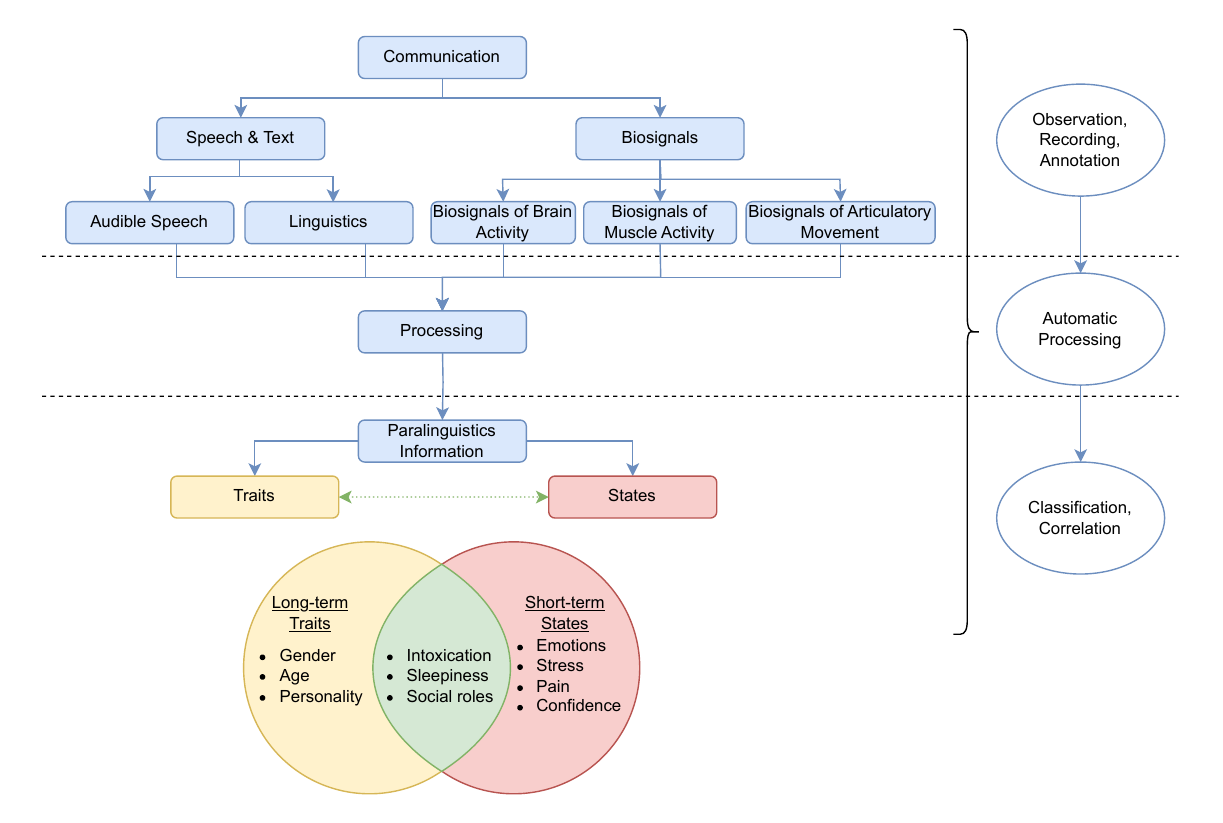}
    \caption{Framework of Computational Paralinguistics using speech and biosignals. During human communication, speech signals, texts, and biosignals can be recorded. Paralinguistic information can then be classified or correlated by processing the audible speech, linguistic content, and biosignals which capture brain activity, muscle activity, and articulatory movement. Paralinguistic information is typically categorized into short-term states (\eg emotion, intention) and long-term traits (\eg personality, gender). Additionally, intermediate states (\eg intoxication, sleepiness) exist and do not fit neatly into this binary distinction.}
    \label{fig:compara}
\end{figure*}

CP aims to automatically recognize paralinguistics information from audible speech, linguistic (\ie written) information, and biosignals. 
This information can pertain to age, gender, sleepiness, politeness, or emotions such as frustration (see Fig. \ref{fig:compara}). Although it focuses on speech signals and biosignals during audibly speaking, it serves as a background topic for SP.

CP can be defined by dividing the term into its two name-giving components\cite{schuller2014computational}.
First, \textit{computational} in this context means that a computer is assisting humans in their work. The work of the computer can range from simple tasks to complete analysis and processing of the data. In contrast to linguistics, paralinguistics is more interested in how something is said than in language structure, grammar, or semantics. This includes prosody, tone of voice, and the selection of particular words among various candidates with similar meanings but distinct connotations~\cite{batliner2020ethics}. Not only is spoken language considered, but also transcribed language. Paralinguistics in text can be represented through connotations, punctuation, capitalization, and ellipses, among other elements.
Second, \textit{paralinguistics} is concerned with the long-term traits, short-term states, and medium-term between traits and states of people, as mentioned in Section~\ref{sec:silentpara}. Long-term traits can also be a cultural characteristic, such as being a native speaker of a first language, in addition to age, gender, and personality. In contrast, short-term states include emotions, or emotion-related states and affects (\eg stress, confidence, and pain). Medium-term between traits and states includes self-induced temporary states (\eg sleepiness or intoxication), structural (behavioural, interactional, and social) signals (\eg roles in dyads and groups), 
and modes (\eg speaking style and voice quality)~\cite{schuller2014computational}.

\subsection{Computational Paralinguistics from Speech}
CP encompasses a broad spectrum of application domains, focusing on the automatic recognition of speaker traits, states, and medium term between the two. 

Regarding long-term traits, applications in multimedia further highlight the versatility of the field: targeted advertising~\cite{batliner2006} and adaptive content delivery can be facilitated by estimating speaker attributes such as age and gender via voice-based interfaces. Insights into how personality traits (\eg the Big Five) manifest in speech enable more naturalistic speech synthesis, thereby supporting the design of socially intelligent systems like empathetic global positioning system navigators~\cite{mohammadi2010voice}.

The applications of detecting short-term states are widespread in computer games, education, health, and many others.
For instance, one line of investigation has examined verbal expressions of students during playing a video game, aiming to infer their level of frustration from their speech~\cite{song2019,song2021}. A related study explored the automatic classification of children’s speech into frustrated, polite, and neutral categories within a gaming context~\cite{yildrim2005}, underscoring the potential of paralinguistic methods to enhance user experience and human-computer interaction. The acoustic differentiation between polite and impolite requests has been consistently observed in~\cite{caballero2018}.
In customer service environments such as call centers, recognizing vocal indicators of anger enables adaptive response strategies designed to improve customer satisfaction~\cite{burkhardt2005,macary2020allosat}. 

Medium-term between traits and states offers another fertile area for analysis. For instance, speech characteristics have been leveraged to assess sleepiness in patients with sleep disorders~\cite{martin2019}, as well as to identify signs of alcohol intoxication~\cite{schuller2011interspeech}. 


\subsubsection{Features in Computational Paralinguistics from Speech}
Acoustic features form the cornerstone of CP, enabling the analysis and inference of paralinguistic information. Early systems in CP were predominantly based on handcrafted features, which offered interpretable and domain-specific insights into speech signals. These included prosodic features (\eg pitch, energy, and duration), spectral characteristics (\eg spectral slope), voice quality indicators (\eg formants), and cepstral descriptors (\eg Mel-Frequency Cepstral Coefficients, \acp{MFCC})~\cite{schuller2004,weninger2011fusing}. Extracted typically at the frame level, these low-level descriptors served as inputs to various statistical learning models.
Over time, research expanded this feature inventory to more comprehensive and standardized representations. Notable examples include 
task-specific sets such as the extended Geneva Minimalistic Acoustic Parameter Set (eGeMAPS) or the broader ComParE2016~\cite{eyben2015geneva} set, which were designed for affective computing and more general voice research applications~\cite{schuller2016interspeech}.
Despite their widespread adoption, handcrafted features are inherently limited in their ability to model the variability and complexity of natural speech. 

With the advent of DL, the focus gradually shifted toward deep representations that could be extracted directly from simple time-frequency representations or raw speech signals. Early efforts employed 
temporal or 
spectrogram-based inputs within convolutional neural networks (CNNs)~\cite{Trigeorgis2016} 
or 
attention-enhanced recurrent models such as \ac{LSTM}-RNNs~\cite{mirsamadi2017,tzirakis2018end2you,zhang2019attention,markitantov2020transfer}. These architectures facilitated the automatic discovery of high-level acoustic patterns without the need for manual feature engineering.
The introduction of self-supervised learning further transformed feature extraction pipelines. 
Self-supervised-learning-based models leverage large volumes of unlabeled audio to learn context-rich embeddings from raw waveforms. These embeddings have demonstrated superior performance across a variety of downstream paralinguistic tasks.
The transition from handcrafted descriptors to deep, self-supervised representations offers improved generalization settings while maintaining or exceeding the performance of traditional approaches~\cite{zhang2024paralbench}, while their interpretability is very limited.

\subsubsection{Machine Learning in Computational Paralinguistics from Speech}
Early computational paralinguistic systems predominantly relied on conventional classifiers, including \acp{SVM}~\cite{chavhan2010speech}, \acp{HMM}~\cite{nwe2003speech}, and \acp{GMM}~\cite{reynolds2002robust}.
Over the past two decades, the evolution of feature representations has correlated with a transition from traditional ML methods to DL architectures. 
Modern architectures employ \acp{DNN} that operate on time-frequency inputs, such as spectrograms, using \acp{CNN}~\cite{keren2016convolutional} and \acp{RNN}~\cite{mishra2024speech}. Two-dimensional CNNs applied to time-frequency representations are typically used to capture local spectral patterns, which are often followed by \ac{LSTM} or Gated Recurrent Unit (GRU) layers to model temporal dependencies~\cite{Trigeorgis2016,mirsamadi2017}.
A major change brought the introduction of Transformer architectures~\cite{vaswani2017attention,wagner2023dawn}, whose self-attention mechanisms significantly improved the modelling of long-range dependencies in speech signals. Building upon this foundation, large self-supervised models have been trained on extensive speech corpora. Encoders from models such as Wav2Vec~\cite{baevski2020wav2vec}, HuBERT~\cite{Hsu21huBERT}, and WavLM~\cite{chen2022wavlm} generate rich representations, offering robust features for a wide array of downstream tasks. These foundation models are typically used either in a frozen state or fine-tuned on paralinguistic data~\cite{Georgios2023}.
In downstream applications, lightweight architectures are often employed. Common strategies include mean temporal pooling over the final encoder layer~\cite{Yang2021Superb}, and learned weighting and averaging across multiple encoder layers~\cite{tsai2022superbsg}.
`Multimodal'
approaches, which integrate audio-derived embeddings with their corresponding textual transcripts, offer an option for improved performance in paralinguistic tasks~\cite{Xu2019LearningAF}. Recent architectures increasingly rely on pre-trained models in both modalities, such as Wav2Vec 2.0 for speech and \ac{BERT} variants or large language models for text~\cite{chen2024_2}, enabling a unified framework for multimodal paralinguistic computations.

\subsubsection{Challenges}
First, the small amount of available corpora is a major limitation of CP. Unlike the datasets for \ac{ASR}, where it is much easier to mix and match datasets to train the data-intensive DL models, the corpora 
use different annotations and recording protocols depending on the task, which limits the amount of data extremely. The annual Computational Paralinguistics Challenge~\cite{schuller2009interspeech,Schuller2010interspeech,schuller2011interspeech,schuller2012interspeech,schuller2013interspeech,schuller2014interspeech,schuller2015interspeech,schuller2016interspeech,schuller2017interspeech,schuller2018interspeech,schuller2019interspeech,schuller2020interspeech,schuller2021interspeech,schuller2022acm,schuller2023acm} addresses this problem by promoting available data sets and thus boosting their development~\cite{vetrab2023}. 

The second challenge is dealing with utterances of different durations since the task of CP is often utterance-level classification or regression. However, standard speech processing features (such as filter banks or \ac{DNN} embeddings) are typically frame-level features. Approaches are needed to transform frame-level features into fixed-size utterance-level features without losing information. This would allow different lengths of audio recordings to be associated with a single label~\cite{vetrab2023aggregation}. 

The third challenge is the lack of less posed and more realistic data~\cite{schuller2013}. Emotions that occur naturally may look different from those that are acted out~\cite{hoque2011acted}, which limits the application of these models in real-world scenarios. However, solving this issue is not easy as collecting and processing realistic data requires a significant amount of effort~\cite{schuller2013}. 

Finally, cross-cultural aspects in paralinguistics are even more challenging~\cite{schuller2013}. Cultural differences can change remarkably the performance of speech emotion recognition~\cite{KAMARUDDIN2012}. Pitch, pitch range, intensity, intensity range, \ac{HNR}, and hesitations and pauses can differ between cultures when expressing politeness~\cite{grawunder2014politeness}. To create intercultural CP systems, data must be collected and combined from various cultural backgrounds.

\begin{table*}[h!]
    \centering
        \caption{Existing studies in computational paralinguistics from biosignals recorded during audible speech. AR: autoregressive, MAV: mean amplutide value, RMS: root mean square, SSC: slope sign change, ZCR: zero crossing rate.
        }
        \scalebox{0.9}{
    \begin{tabular}{p{2.5cm}p{0.6cm}p{1.5cm}p{4cm}p{3.4cm}p{2.4cm}p{2.4cm}}
    \toprule    
         \textbf{Paper}& \textbf{Year} & \textbf{Input} & \textbf{Feature} &\textbf{Method} & \textbf{Data} & \textbf{Output} \\  
    \midrule
        Nakamura \etal\cite{Nakamura2011} & 2011 & \ac{EMG} & TD15 & \ac{SVM}, \ac{GMM} & Self-recorded data  &Voiced/unvoiced, \ac{F0}\\

        Armas \etal\cite{winston2014} & 2014 & \ac{EMG}, Respiratory Trace & \ac{EMG}: RMS, MAV, AR model coefficients, waveform length, Respiratory Trace: max. derivative, min. derivative, max. trace, min. trace, integral of the trace  & \ac{SVM} & Self-recorded data  & \ac{F0}, voicing state\\

        Diener \etal\cite{9003804} & 2019 & \ac{EMG} & TD15 & \ac{FNN} & Self-recorded data  & \ac{F0}, speech\\



        Vojtech \etal\cite{Vojtech2022} & 2022 & EMG & (Beta) coherence, central frequency, cross-correlation, wavelet coefﬁcients (maximum/peak/mean/variance), maximum (peak) frequency, mean absolute value, (mean/median) frequency, mean power density, \ac{MFCC}, power density wavelength, RMS, SSC, spectral moments, variance, waveform length, ZCR & \ac{FDNN} & Self-recorded data  & \ac{F0}, intensity contour\\
    \midrule    
        Hui \etal\cite{Hui2015} & 2015 & \ac{EGG} & Pitch & Statistical Model: ANOVA &  Self-recorded data & Emotion\\

        Pravena \etal\cite{pravena2017development} & 2017 & \ac{EGG} & \ac{SoE}, \ac{F0} & \ac{GMM} &  EmoDB\cite{burkhardt2005database}, self-recorded data & Emotion\\

        Pravena \etal\cite{pravena2017significance} & 2017 & \ac{EGG} & \ac{SoE}, delta \ac{SoE}, delta-delta \ac{SoE}, \ac{F0}, delta \ac{F0}, delta-delta \ac{F0}, \ac{MFCC} & \ac{GMM} &  EmoDB\cite{burkhardt2005database}, Tamil-Db\cite{pravena2017development} & Emotion\\
    \midrule
        Liu \etal\cite{Liu2016} & 2016 & \ac{EMA} & \ac{EMA}, Power, Voiced/unvoiced flags, \ac{F0} & \ac{GMM}, \ac{FDNN}, \ac{RNN}, \ac{LSTM}-RNN &  MNGU0~\cite{richmond11_interspeech}  & Speech, power, Voiced/unvoiced, \ac{F0}\\

        Zhao \etal\cite{Zhao2018} & 2018 & \ac{EMA} &\ac{EMA}, delta \ac{EMA}, delta-delta \ac{EMA}, and \ac{MFCC} & \ac{FDNN}, \ac{LSTM}-RNN & MNGU0~\cite{richmond11_interspeech}  & Voiced/unvoiced, \ac{F0}\\

        Stone \etal\cite{9053231} & 2020 & \ac{EMA} & \ac{EMA}, delta \ac{EMA}, delta-delta \ac{EMA} & \ac{SVM}, \ac{KRR}, \ac{FDNN} & MNGU0~\cite{richmond11_interspeech}  & Voiced/unvoiced/silent, \ac{F0}\\
    \midrule
        Grosz \etal\cite{Grosz2018} & 2018 & Ultrasound & Raw data & \ac{FDNN} & Self-recorded data  &  Voiced/unvoiced, \ac{F0}\\
        Dai \etal\cite{Dai2021} & 2021 & Ultrasound & Raw data & \ac{FDNN} & Self-recorded data  & Voiced/unvoiced, \ac{F0}\\
        
    \bottomrule
    \end{tabular}
    }
    \label{tab:com_para}
\end{table*}

\subsection{Computational Paralinguistics from Biosignals}
\label{sec:cp_biosignal}
The aim of CP from biosignals is to analyze and interpret non-verbal aspects of communication. This involves extracting meaningful patterns from biosignals of brain activity, muscle activity, and articulatory movement, to understand how these signals correlate with paralinguistic cues. However, there has been limited research on detecting paralinguistic states and traits from biosignals during speech. Most existing work focuses on identifying paralinguistic cues while individuals are engaged in other activities, such as listening to music or watching videos. 

In related studies about detecting paralinguistic cues from biosignals during speech, most works focused on audible speech rather than silent speech. The detected information mainly includes prosodic features, such as \ac{F0}, pitch, energy, duration, and frame-level features, which have been found to correlate with paralinguistic information, such as emotions~\cite{HASHEM2023102974}. The state-of-the-art studies using \ac{EGG}, \ac{EMA}, \ac{EMG}, and ultrasound images are presented in Table~\ref{tab:com_para}.
Regarding each type of biosignals, we will generally introduce detecting paralinguistic cues from bisignals, and then discuss relevant detection works of biosignals recorded during audible speech.

\subsubsection{Biosignals of Brain Activity.}
Biosignals which capture brain activity have been applied to detect a range of paralinguistic states and traits. For instance, such biosignals have shown the potential to recognize discrete emotional states, such as happiness, sadness, and anger~\cite{schaaff2009towards,schaaff2009towardseeg,Bhatti2016-HER,Spezialetti2018-TEB,Houssein2022-HER,Zhao2017-EAF}.
Cognitive states have also been distinguished via biosignals of brain activity. One study using \ac{EEG} alone reported classification of states such as relaxation, neutral, and concentration~\cite{Bird2018-ASO}, and another study integrating multiple biosignals, including pulse of blood volume and eye movements, recognized cognitive states such as frustration~\cite{Akira2015-DOC}. Two studies further combined biosignals of brain activity with other physiological measures to evaluate personality traits~\cite{Zhao2017-EAF, Wache2014-TSL}. To the best of the authors' knowledge, there is no study of paralinguistics on biosignals of brain activities during audible speech.

\subsubsection{Biosignals of Muscle Activity.}
Biosignals of muscle activity have been applied to recognize a range of paralinguistic emotional states. \ac{EMG} signals, in combination with other signals such as \ac{ECG}, electrodermal activity, and respiratory signals, have been used to recognize discrete emotional states~\cite{Kim2008-ERB,Maaoui2010-ERT}.
Furthermore, facial \ac{EMG} signals from the corrugator supercilii and zygomaticus major muscles have been used to distinguish valence and arousal states in young and senior adults~\cite{Tan2016-ROI}. A combination of electrodermal activity and facial \ac{EMG} signals has also been employed to classify neutral, positive, negative, and mixed emotions~\cite{Broek2010-AMI}.

\textbf{Biosignals of muscle activity during speech. }
As shown in Table~\ref{tab:com_para}, Nakamura \etal\cite{Nakamura2011} presented an \ac{F0} estimation method from EMG signals during audible speech. Additionally, voiced/unvoiced frames were detected using an SVM, and \ac{F0} was predicted using a \ac{GMM}-based model. 
For the same two tasks, Armas \etal~\cite{winston2014} collected EMG and respiratory trace signals to estimate \ac{F0} and voicing state. Specifically, three types of EMG features and five types of respiratory trace features were extracted, and an SVM classifier was used to predict \ac{F0} and voicing state. 
Diener \etal~\cite{9003804} proposed an approach to predict \ac{F0} for EMG-to-speech conversion. 
Time domain-based TD15 features~\cite{jou2006towards} were extracted from EMG signals, and fed into an \ac{FDNN} model for \ac{F0} prediction. 
Vojtech \etal\cite{Vojtech2022} used \ac{FDNN} for both \ac{F0} and intensity contour prediction based on \ac{EMG} features, including time-domain, frequency-domain, and cepstral-domain features.

\subsubsection{Biosignals of Articulatory Movement.}
Research advances in articulatory research have underscored the role of biosignals in the detection of paralinguistic states and traits. 
An early work by Erickson et al.~\cite{Erickson2000-ACO} employed \ac{EMA} to capture jaw and tongue dorsum adjustments across emotional contexts. Their results showed that articulatory configurations, along with acoustic shifts in \ac{F0} and formant frequencies, provide reliable indicators of emotion. Notably, spontaneous emotional speech was marked by stronger lingual gestures. This is particularly indicated by higher and more fronted tongue positions~\cite{Erickson2004-SAM}. 
Kim et al.~\cite{Kim2010-ASO} further demonstrated that anger and happiness could be distinguished by their unique weighting of articulatory and prosodic features, while their subsequent work emphasized that estimated articulatory trajectories convey complementary paralinguistic cues and support emotion classification. Complementing these studies, Zhang et al.~\cite{Zhang2023-ASO} found that integrating articulatory cues improved emotion recognition in Mandarin across multiple categories and intensity levels, emphasizing the universal relevance of articulatory biosignals in emotion recognition.

\textbf{Biosignals of articulatory activity during speech. }
In an EGG-based study~\cite{Hui2015}, pitch derived from EGG has been found to be related to emotional states, including neutral, happy, and sad. However, only a weak interaction between pitch and personality was reported. The work in~\cite{pravena2017development,pravena2017significance} investigated features, especially \ac{F0}- and \ac{SoE}-based features, for emotion recognition using \ac{GMM}. 

Zhao \etal\cite{Zhao2018} predicted voiced/unvoiced flags and \ac{F0} from \ac{EMA} in the MNGU0 database~\cite{richmond11_interspeech}. 
Similarly, Stone \etal~\cite{9053231} predicted voiced/unvoiced/silent segments of \ac{EMA} data using multiple models, including \ac{SVM}, \ac{KRR}, \ac{FDNN}. 
Liu \etal\cite{Liu2016} proposed an \ac{ATS} method using \ac{GMM}, \acp{FDNN}, and LSTM-RNNs based on \ac{EMA} in the MNGU0 database~\cite{richmond11_interspeech}. In addition to spectral features, excitation features such as power, voiced/unvoiced flags, and \ac{F0} related to paralinguistics were predicted to synthesize speech. 

Based on ultrasound images, Grosz \etal~\cite{Grosz2018} estimated voicing of the actual frame and predicted \ac{F0} of the voiced frames using \acp{FDNN}. 
In~\cite{Dai2021}, \acp{FDNN} were also used to detect voiced/unvoiced segments and for \ac{F0} prediction. 

\section{Affective Speech Synthesis}\label{sec:emotional_voice_conversion}

The affective speech synthesis in this work aims
to integrate paralinguistic information into speech. 
Affective speech synthesis can improve human-human and human-machine conversations~\cite{tits2020ice,crumpton2016survey}. It can be integrated into robots which play many roles, such as companions, tutors, caregivers, colleagues, etc. 
Furthermore, affective speech synthesis can be promising to leverage
speech to be more expressive for individuals who suffer from laryngectomy~\cite{NAKAMURA2012}. It can be developed in an indirect way, which converts synthesized speech into another paralinguistic style using \textit{affective speech conversion}. 
Alternatively, it can be processed 
in a direct way that integrates paralinguistic information into biosignal-to-speech conversion. As there is a lack of research in directly converting biosignals into affective speech, we will introduce affective speech conversion used in the indirect method.

Affective speech conversion is a technique that converts the state of a speaker's utterance while preserving the linguistic information and identity of the speaker. 
While speech synthesis technology has advanced towards human-like naturalness, it cannot still convey human-like affective states. Therefore, affective speech conversion has a huge potential for applications in human-machine interaction. 
We will introduce affective speech conversion models with parallel and non-parallel data. 

\begin{figure}[ht]
    \centering
    \includegraphics[width=0.49\textwidth]{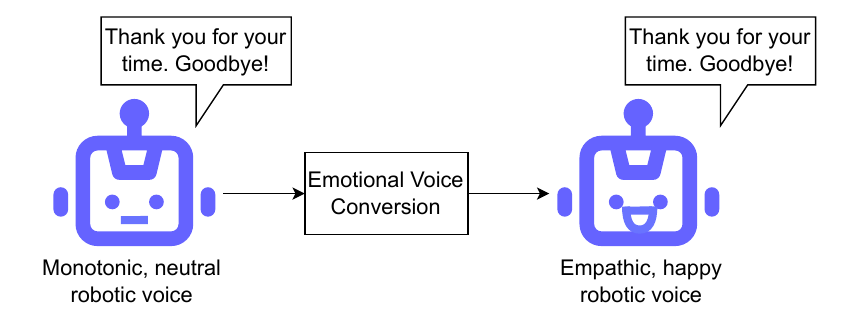}
    \caption{Application of affective speech conversion for human-machine-interaction.}
    \label{fig:evc}
\end{figure}

\subsection{Affective Speech Conversion with Parallel Data}
In affective speech conversion with parallel data, pairs of speech recordings with the same linguistic content are used to learn the mappings between different expressions. The usual processing steps include feature extraction, frame alignment, and feature mapping~\cite{Zhou2022}. Given pairs of speech, directly comparing the source- and target-speech can be efficient during training.
Nevertheless, since parallel data is often recorded for a small number of affective states and participants in a laboratory environment, there are difficulties to adapt trained models to unseen states and speakers. Data recording is also costly and time-consuming in lab environments. These points lead to difficulties in applying parallel data-based models for real-world applications.

\begin{figure}
    \centering
    \includegraphics[width=\linewidth]{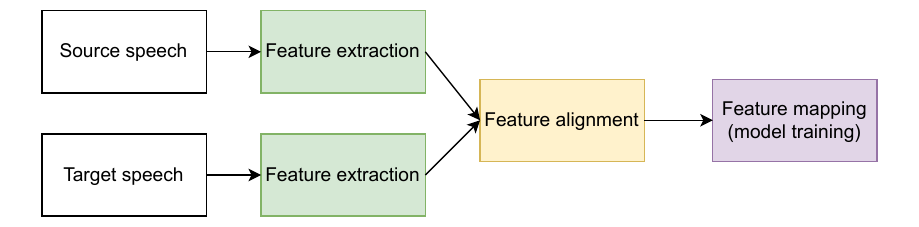}
    \caption{The framework of affective speech conversion with parallel data.}
    \label{fig:voiceconvert}
\end{figure}

\textbf{Feature Extraction and Alignment.}
Spectral features and prosody features are generally used during feature extraction. \ac{MFCC}, \ac{LPCC}, and \ac{LSF} are usually used as spectral features~\cite{Zhou2022,sisman2020overview}. Prosodic features include elements such as \ac{F0}, jitter, speaking rate, and energy contour. Then, the features from the source speech and target speech are aligned at the frame level.

\textbf{Feature Mapping.}
Statistical models can achieve feature mapping with aligned paired utterances. \ac{GMM}-based conversion models modify prosodic and voice quality information while keeping linguistic information unchanged~\cite{aihara2012gmm}. Other statistical models such as \ac{PLSR}\cite{helander2010voice} have been usedin the last decades. Additionally, non-negative matrix factorization\cite{lee2000algorithms} approaches were designed to address the over-smoothing problem.

In recent years, neural network approaches have become state-of-the-art for affective speech conversion with parallel data. For example, the \ac{DNN} in~\cite{chen2014voice} constructs a global non-linear mapping relationship between the spectral envelopes of two speakers. Deep belief networks can change the spectrum and the prosody for the emotional voice at the same time~\cite{Luo2016}. The \ac{biLSTM}-RNNs do not model only the frame-wised relationship between the source and the target voice but also the long-range context-dependencies in the acoustic trajectory~\cite{sun2015voice, XUE2018}. However, \acp{DNN} are limited for spectrum and prosody transformation because they need large amounts of parallel data~\cite{sisman2020overview}.

\subsection{Affective Speech Conversion with Non-parallel Data}
Given that recording parallel data is expensive, it is essential to convert affective speech with non-parallel data. Instead, one can
utilize separate sets of recordings for each emotional state, where the utterances differ across the states. Using non-parallel data makes it possible to record more data, while models and training are more complex.
There are three types of approaches: translation models, disentanglement between prosody and linguistic content, and multi-task learning models.

\textbf{Translation Models.}
Translation models between the source and target domains have been studied for non-parallel data, including models like \acp{GAN}~\cite{rizos2020stargan} and CycleGANs~\cite{kaneko2018cyclegan,zhou2020transforming}. While CycleGAN-VC~\cite{kaneko2018cyclegan} and CycleGANVC2~\cite{kaneko2019cycleganvc2} are still based on \acp{MFCC}, logarithmic fundamental frequency and aperiodicity, CycleGANVC3\cite{kaneko2020cyclegan} uses Mel spectrograms as feature input~\cite{Chun2023}. Li~\cite{li2021starganv2} proposed an adapted StarGANv2 for voice conversion using a combination of adversarial source classifier loss and perceptual loss. It produces natural-sounding voices close to \ac{TTS}-based voice conversion models. Diffusion models, such as EmoConv-Diff~\cite{Prabhu2024} and DiffHier-VC~\cite{choi2023diff}, have been successfully applied. A new promising architecture is modified CycleGANs incorporating the embedding of a stride-denoised multimodal diffusion model called DiffGAN-VC. DiffGAN-VC overcomes the problem of expensive and time-consuming computation for diffusion models with \acp{GAN} and uses a large-step denoising operation. So far, it has only been evaluated successfully on speaker identity voice conversion but can be extended to affective speech conversion~\cite{zhang2023voice}.

\textbf{Disentanglement between prosody and linguistic content.}
This approach type disentangles the linguistic and non-linguistic information hidden in the source and target voice respectively, and then combines
them for converting speech~\cite{zhao2022disentangling}.
An auto-encoder~\cite{Qian2020} can be used to separate speech style and linguistic content. \acp{VAE-GAN} can extract content and emotion-related representations and recombine them in the conversion stage~\cite{cao2020nonparallel}. The \ac{VAW-GAN} enables speaker-independent emotional voice conversion by mapping of spectrum and prosody representations\cite{zhou2020converting}. The StarGANv2 architecture was recently extended by utilizing dual encoders to learn the speaker and emotion style embeddings separately, using unseen speaker-emotion pairs~\cite{Shah2023}.

\textbf{Multi-task learning.}
Multi-task learning with \ac{TTS}, \ac{SER}, or \ac{ASR} systems has been studied for affective voice conversion~\cite{Zhou2022}. As datasets in these tasks have large amounts of samples, they can be used for model pre-training. For example, Emovox\cite{zhou2022emotion} is a sequence-to-sequence framework with a parallel WaveGAN vocoder that converts voice by separate embeddings for the linguistic content, emotion, and emotion intensity. It uses a \ac{TTS} corpus to pre-train the emotion encoder as a speaker-style encoder. 
In~\cite{zhou2021}, an SER descriptor is pre-trained to transfer emotional style that generates deep emotional feature representations for voice conversion. Results of \ac{ASR} have been exploited successfully to improve the preserving of linguistic contents in StarGAN-VC~\cite{Sakamoto_2021}.

\subsection{Challenges}
The primary challenge in the field of affective speech synthesis is the amount of appropriate public datasets. The existing datasets are either small or of poor quality, which hinders further research in this promising field and limits the performance improvement~\cite{yang2022overview}. Moreover, many databases have limited speaker, language, and lexical variability, further complicating the training process of robust \ac{EVC} models. As with CP, cultural differences in the expression of emotions are therefore challenging\cite{Zhou2022}.

Furthermore, accents in the training data also cause a problem, as this is paralinguistic information and should not be transferred during the conversion of a speaker without an accent but should be transferred for a speaker with an accent. The same applies to paralinguistic information such as age and gender~\cite{Zhou2022}.

Moreover, relevant models are typically trained on high-quality acted-out speech data, therefore, the models are sensitive to noise and real-world variability. Models trained on acted-out speech may generate stereotypical representations of emotions. Therefore, more models trained on in-the-wild data are needed~\cite{Prabhu2024}.  

The final challenge is the lack of objective evaluation methods. Many evaluation methods are not intuitive and understandable enough. The available metrics only indicate how similar the converted speech is to the target, yet, this does not guarantee that humans will perceive them as being close~\cite{yang2022overview}.


\begin{table*}[h!]
    \centering
        \caption{Existing studies in SP. MAV: mean amplutide value, RMS: root mean square, SSC: slope sign change, ZCR: zero crossing rate.}
        \scalebox{0.95}{
    \begin{tabular}{p{1cm}p{0.6cm}p{1cm}p{4cm}p{2.4cm}p{2.4cm}p{1.5cm}p{2cm}}
    \toprule
         \textbf{Paper}& \textbf{Year} & \textbf{Input} & \textbf{Feature} &\textbf{Method} & \textbf{Data} & \textbf{Mode} & \textbf{Output} \\  
    \midrule
        Diener \etal\cite{diener2020towards} & 2020 & \ac{EMG}& MAV, RMS, sum absolute values, variance, simple square integral, waveform length, average amplitude change, ZCR, SSC, median/weighted-mean frequency, recurrent autoencoder-based deep features & \ac{LDA}, Random Forest, \ac{LSTM}-RNN, \ac{SVM} & EMG-UKA~\cite{wand14b_interspeech,ELRA-S0390} & Audible, Silent, Whispered & speaker identity, speaking mode (audible, silent, whispered)\\


        Vojtech \etal\cite{vojtech2021surface} & 2021 & EMG & \ac{MFCC} & \ac{HMM}, \ac{GMM}, decision tree, maximum likelihood linear regression, \ac{FNN} & Self-recorded data & Silent & Text, speech, stress \\

        Wairagkar \etal\cite{wairagkar2025instantaneous} & 2025 & Intracortical signals & Spike-band power features & Encoder (Transformer), vocoder (LPCNet) & Self-recorded data & Silent& Speech, speaking speed, question intonation, pitch\\
        
    \bottomrule
    \end{tabular}
    }
    \label{tab:silent_para}
\end{table*}

\section{Current Research and Open Challenges in Silent Paralinguistics}\label{sec:CurrentResearch}
\subsection{Currect Research} \label{sec:current_research}

To the best of the authors' knowledge, there are only a few studies in the field of SP. 
The state-of-the-art studies mainly used intracortical signals and \ac{EMG} signals, which are presented in Table~\ref{tab:silent_para}.
It has been demonstrated that, paralinguistic features can be detected from intracortical signals when a participant silently speaks a sentence without vocalizing~\cite{wairagkar2025instantaneous}. The detected paralinguistic features included speaking speed, question intonation, and pitch. Therefore, the participant can modulate his/her synthesized voice. The ground truth of this work was based on synthesized speech from given texts using a \ac{TTS} algorithm.

\ac{EMG} signals are related to both articulatory muscle activity and facial expressions, which are associated with paralinguistics~\cite{botelho2020toward}. 
There have been studies focusing on detecting paralinguistic information from \ac{EMG} during silent speech.
Diener \etal~\cite{diener2020towards} detected speaker identity and speaking mode (audible, silent, and whispered) from \ac{EMG} signals, and synthsized speech for SP purposes, revealing that SP is feasible. Specifically, the work~\cite{diener2020towards} utilized EMG-UKA parallel EMG-Speech corpus~\cite{wand14b_interspeech, ELRA-S0390}, which consists of 8 speakers and 63 sessions. 
In another study~\cite{vojtech2021surface}, phonetics and stress were detected from \ac{EMG} signals during silent speech, and were then integrated into a \ac{TTS} model for generating audible speech.


\subsection{Open Challenges and Future Directions}
In the following, we will discuss the challenges and future work from multiple perspectives, including paralingsuistics in silent speech, datasets and biosignals.

\subsubsection{Paralinguistics in Silent Speech}
Although prosodic features can be predicted from biosignals such as \ac{EMA}, \ac{EMG}, and ultrasound (see Section~\ref{sec:cp_biosignal}), only a few proposed models have been evaluated in silent speech (see the aforementioned studies in Section~\ref{sec:current_research}). 
Grosz \etal~\cite{Grosz2018} proposed a future work to assess their system on ultrasound images during silent speech; however, to the authors' knowledge, these \ac{F0} prediction algorithms have not been evaluated in the context of silent speech. In fact, this can be challenging, since ground truth for \ac{F0} is not available for silent speech, and the retention of prosody information during silent speech remains unclear. Therefore, SP can benefit from speech synthesis models integrated with paralinguistic information which can be annotated in silent speech, such as emotions. One can work with the following steps: obtaining the ground truth of paralinguistic information from modalities such as facial images/videos, training models to predict acoustic features and paralinguistic information, and finally incorporating the predicted features into a speech synthesis system.

\subsubsection{Databases}



A common limitation of existing datasets is the number of speakers. The MNGU0~\cite{richmond11_interspeech} dataset and the study by Grosz \etal\cite{Grosz2018} each include recordings from only one speaker and single language. In \acp{SSI}, it is essential to train a generalized model for synthesizing speech from multiple speakers and multiple languages. For SP, the detection of speaker traits and states also require multiple speakers' data for training powerful models, thereby finally synthesizing expressive speech with a high-level personalization. 
There is also a need for more comprehensive datasets that encompass multiple speakers' cross-culture data with multiple languages. Alternatively, \ac{TTS} technologies can help to generate speech of multiple speakers for model training, and provide diverse choices of voice for voiceless people~\cite{scheck2025diffmv}.

Furthermore, among the existing datasets, only the EMG-UKA corpus\cite{wand14b_interspeech,ELRA-S0390} and the other two studies~\cite{vojtech2021surface,wairagkar2025instantaneous} (one is EMG corpus and the other is intracortical-signal corpus) include biosignal data during silent speech. It is also essential to develop other potential data modalities that can be captured in a comfortable and convenient way, such as ultrasound reflected by mobile devices~\cite{Su696,gao2020echowhisper}.

Additionally, a common drawback of these datasets is that they predominantly involve reading tasks, such as reading newspaper sentences. Natural conversation, however, includes more varied and unrestricted communication, with greater emotional and prosodic content. Therefore, it is crucial to develop datasets that incorporate emotional expressions and more natural conversational contexts.

Finally, future studies need to be conducted with actual laryngectomy patients~\cite{winston2014}.
There are differences in the neck musculature and articulatory movements of laryngectomy patients that make it difficult to predict how well current approaches extend to them~\cite{winston2014,Grosz2018,Dai2021}.

\subsubsection{Biosignals}
\textbf{Electroglottography.}
\ac{SoE} is an \ac{EGG}-feature that has potential to improve emotion recognition~\cite{pravena2017significance}. One can investigate how it performs in SP tasks.
EGG-based \ac{F0} and \ac{SoE} features are showing different behaviors in emotion recognition compared to speech-based \ac{F0} and \ac{SoE}. The connection between these two modalities can be studied~\cite{pravena2017significance}.
Emotion recognition with \ac{EGG} can be extended to more emotional classes or dimensions (such as arousal and valence) than currently studied discrete emotions~\cite{Hui2015,pravena2017significance}.

\textbf{Electromagnetic Articulography.}
Open challenges in \ac{EMA} address if pitch accents are recognizable from the supraglottal articulation and can subsequently be correctly synthesized~\cite{9053231}.
Analyzing more expressive voice intonation is also necessary because the datasets that are currently available only include read speech that is neutral in emotion~\cite{9053231}.
Utilizing different ML approaches~\cite{Zhao2018,Liu2016} and more articulatory input (such as fMRI images) holds the potential to enhance \ac{EMA}-to-\ac{F0} prediction~\cite{Liu2016} and provide steps into the direction of articulatory-controllable speech synthesis~\cite{Zhao2018,Liu2016}.

\textbf{Electromyography.}
The primary challenges in predicting voice \ac{F0} and intensity from \ac{EMG} activity are improving multi-speaker models' accuracy. This is because existing techniques rely on single-speaker models which are impractical for voiceless people~\cite{Vojtech2022}. 
Furthermore, different electrode positions could be investigated for better capture of vocal folds information\cite{Nakamura2011}. Future study might develop an automated method of determining optimal electrode placement for every new patient~\cite{winston2014}.
Moreover, to provide a comprehensive prosodic analysis, future work should include normalization techniques for microphone distance variability and expand model capabilities to include timing and voice quality attributes~\cite{Vojtech2022}.

\textbf{Ultrasound.}
Several traditional discontinuous speech-based \ac{F0} estimation algorithms can achieve \ac{F0} prediction from ultrasound tongue images. It is necessary to test how these algorithms trained on single speaker's data work for other speakers or with real silent articulation~\cite{Dai2021}.
Future ultrasound-based SSI systems will inherently be personalized to multiple speaker embeddings. 

\section{Conclusion}
This work defined the research area of silent paralinguistics and highlighted its importance. Silent paralinguistics aims to detect paralinguistic information from silent biosignals, and integrate it into generated text and speech from biosignals. This research area involves multiple procedures using current state-of-the-art technologies, from paralinguitic information perception, to biosignal-based speech recognition and synthesis. The work introduced existing technologies and presented a survey on existing related works. Finally, open challenges were discussed and potential future work was pointed out. This work is expected to shed light on the area of silent paralinguistics, and to promote relevant studies in the speech and biosignal community.

\ifCLASSOPTIONcaptionsoff
  \newpage
\fi

\bibliographystyle{IEEEtran} 
\bibliography{bib/bare_jrnl}

\begin{IEEEbiography}[{\includegraphics[width=1in,height=1.25in,clip,keepaspectratio]{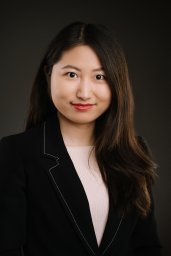}}]{Zhao Ren} (S'19--M'22) is currently a postdoctoral researcher at the University of Bremen, Germany, for silent speech interfaces and human-computer interaction.
She received her bachelor and master degree in Computer Science and Technology from the Northwestern Polytechnical University in China, in 2013 and 2017. She pursued her doctoral degree in University of Augsburg, Germany, in 2017-2022. She then worked as a researcher and project coordinater at L3S Research Center, Leibniz Univerisity Hannover, Germany, for AI applications in healthcare and medicine. She served as guest editors in journals, such as IEEE J-BHI and multiple Frontiers journals, and as session chairs in ICASSP and EMBC.
She regularly reviews top-tier journals such as IEEE TAFFC and IEEE TMM, and conferences such as AAAI, INTERSPEECH, and ICASSP. 
Her research interests mainly lie in computer audition, computational paralinguistics, silent paralinguistics, and their applications in health care and wellbeing. She has (co-)authored more than 70 publications in peer-reviewed book chapters, journals, and conference proceedings, which have received 2.9k+ citations (h-index 27).
\end{IEEEbiography}

\begin{IEEEbiography}[{\includegraphics[width=1in,height=1.25in,clip,keepaspectratio]{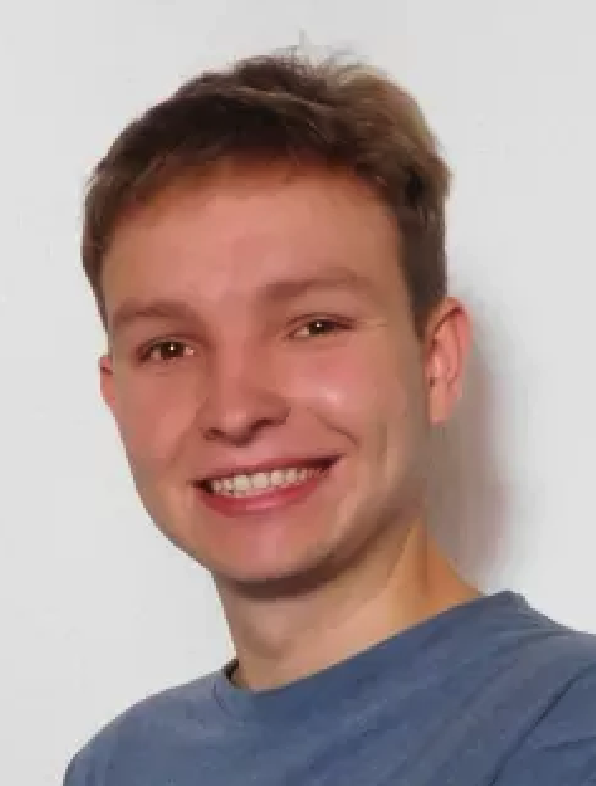}}]{Simon Pistrosch} received his M.Sc. degree in Computer Science from the University of Augsburg in 2022. He is currently a Research Assistant and Ph.D. candidate at the Chair of Health Informatics (CHI), Technical University of Munich (TUM), and a member of the Munich Center for Machine Learning (MCML). His research interests include biosignal and audio processing, emotion recognition, and silent paralinguistics.
\end{IEEEbiography}

\begin{IEEEbiography}[{\includegraphics[width=1in,height=1.25in,clip,keepaspectratio]{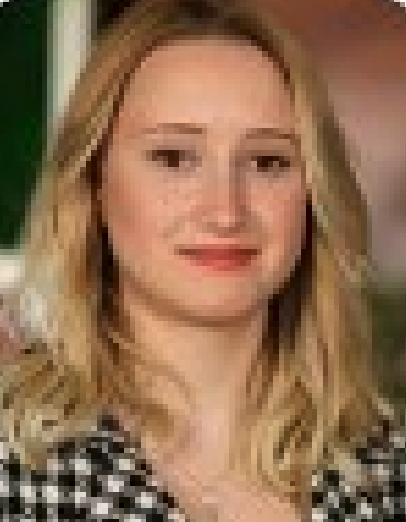}}]{Buket Coşkun} received her B.Sc. degrees in Biomedical Engineering and Electrical and Electronics Engineering from Yeditepe University, Istanbul, in 2021. She worked as a researcher at Yeditepe University from 2021 to 2024, focusing on emotion recognition using physiological signals. She also worked as a researcher at the Cognitive Systems Lab, University of Bremen. Her research interests include biosignal processing, emotion recognition, and silent paralinguistics.
\end{IEEEbiography}

\begin{IEEEbiography}[{\includegraphics[width=1in,height=1.25in,clip,keepaspectratio]{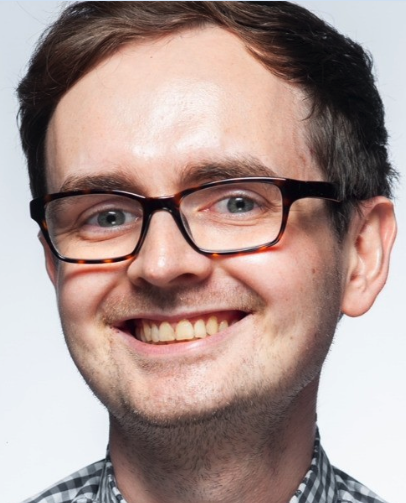}}]{Kevin Scheck} received his M.Sc. degree in computer science from the University of Bremen, Germany, in 2020. He is currently working as a doctoral researcher at the Cognitive Systems Lab, University of Bremen. His research interests include biosignal processing, speech synthesis, and silent speech interfaces. He was co-organizer of three occurrences of the special session ``Biosignal-enabled Spoken Communication" at Interspeech.
\end{IEEEbiography}

\begin{IEEEbiography}[{\includegraphics[width=1in,height=1.25in,clip,keepaspectratio]{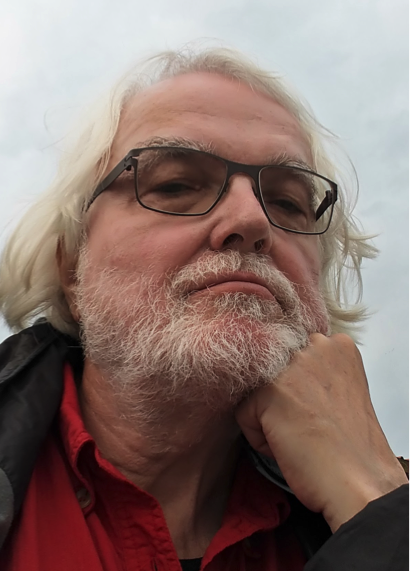}}]{Anton Batliner} received his doctoral degree
in Phonetics in 1978 at LMU Munich. He is now with the CHI - Chair of Health Informatics, Technical University of Munich (TUM), Munich, Germany. He is co-editor/author of
two books and author/co-author of more than 300 technical articles, with an h-index of 55+
and 16000+ citations. His main research interests are all (cross-linguistic) aspects of prosody and (computational) paralinguistics.
\end{IEEEbiography}

\begin{IEEEbiography}[{\includegraphics[width=1in,height=1.25in,clip,keepaspectratio]{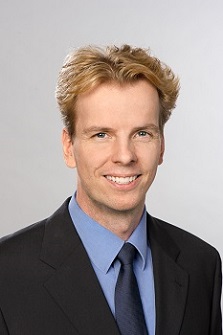}}]{Bj\"orn W. Schuller} (M'06--SM'15--F'18) received his diploma in 1999, his doctoral degree for his study on automatic speech and emotion recognition in 2006, and his habilitation and Adjunct Teaching Professorship in the subject area of signal processing and machine intelligence in 2012, all in electrical engineering and information technology from Technische Universit\"at M\"unchen (TUM), Germany. He is a Full Professor and Head of CHI - Chair of Health Informatics, Technical University of Munich (TUM), Munich, Germany, and a Professor of Artificial Intelligence heading GLAM -- the Group on Language, Audio \& Music, Department of Computing at the Imperial College London in London, UK. Dr.\,Schuller was an elected member of the IEEE Speech and Language Processing Technical Committee, Editor in Chief of the IEEE Transactions on Affective Computing, and is President-emeritus of the AAAC, Fellow of the IEEE, and Senior Member of the ACM. He (co-)authored 7 books and more than 1.7k+ publications in peer reviewed books, journals, and conference proceedings leading to more than 76k+ citations (h-index 121).
\end{IEEEbiography}

\begin{IEEEbiography}[{\includegraphics[width=1in,height=1.25in,clip,keepaspectratio]{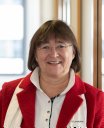}}]{Tanja Schultz} (Fellow, IEEE) received the Diploma and Ph.D. degrees in informatics from the University of Karlsruhe, Germany. She spent over 20 years as a Researcher and an Adjunct Research Professor at Carnegie Mellon University, Pittsburgh, PA, USA. Since 2015, she has been a Professor of cognitive systems at the University of Bremen, Germany. In 2007, she founded the Cognitive Systems Laboratory, where she and her team combine machine learning methods with innovations in biosignal processing to create biosignal-adaptive cognitive systems. She is a fellow of ISCA, in 2016; EASA, in 2017; IEEE, in 2020; and AAIA, in 2021. She received several awards for her work. Currently, she leads the University’s high-profile area “Minds, Media, Machines,” is a speaker of the DFG Research Unit Lifespan AI, and a co-speaker of two research training groups. Recently, she established the international Master’s Program on artificial intelligence and intelligent systems.
\end{IEEEbiography}

\end{document}